\documentclass[12pt,letterpaper]{article}
\usepackage{amsmath}
\usepackage{amssymb}
\usepackage{amsfonts}
\usepackage{amsthm}
\usepackage{mathtools}
\usepackage{geometry}
\usepackage{setspace}
\usepackage{natbib}
\usepackage{mathpazo}
\usepackage[utf8]{inputenc}
\usepackage[T1]{fontenc}
\usepackage{booktabs}
\usepackage{xcolor}
\usepackage{microtype}
\usepackage{tikz}
\usetikzlibrary{arrows.meta,positioning,calc,shapes.geometric}
\usepackage{xurl}
\usepackage{hyperref}
\usepackage{needspace}

\hypersetup{colorlinks,linkcolor=blue!60!black,citecolor=blue!60!black,urlcolor=magenta}

\newtheorem{theorem}{Theorem}
\newtheorem{lemma}{Lemma}
\newtheorem{proposition}{Proposition}
\newtheorem{corollary}{Corollary}
\theoremstyle{definition}
\newtheorem{definition}{Definition}
\newtheorem{assumption}{Assumption}
\renewcommand{\theassumption}{%
  \ifcase\value{assumption}\or 1a\or 2\or 3\or 1b\else\arabic{assumption}\fi}
\theoremstyle{remark}

\allowdisplaybreaks
\begin{document}

\title{\textbf{Who Keeps a Bubble Alive?}\\[4pt]
{\large Skin in the Game and the Survival of Rational Bubbles}
\thanks{%
Feng Dong acknowledges financial support from the National Natural Science Foundation of China (72525004). Siqing Wang acknowledges financial support from the National Natural Science Foundation of China (72503273). Fan Wu acknowledges financial support from the National Natural Science Foundation of China (72603022). The usual disclaimer applies.}}
\author{Feng Dong\thanks{%
School of Economics and Management, Tsinghua University; Email:
dongfeng@sem.tsinghua.edu.cn.} \and Siqing Wang\thanks{%
School of Economics, Central University of Finance and Economics; Email:
wangsq\_cufe@cufe.edu.cn.} \and Fan Wu\thanks{%
HSBC Business School, Peking University; Email:
fanwu@phbs.pku.edu.cn.  \texttt{\ }}}
\date{\today}
\maketitle

\begin{abstract}
We study how retained ownership supports the value of an asset that pays
no dividends and is held only for resale. A sponsor chooses hidden
maintenance that determines market survival, bears its full cost, and
captures only the benefit accruing to its retained tokens. A positive
stationary equilibrium exists if and only if retention meets an
endogenous minimum. Under additional curvature restrictions, retention
above this threshold supports two positive stationary equilibria.
Conditional on survival, prices below the upper equilibrium converge to
the lower one while the upper price is the largest feasible initial valuation.
Greater retention expands the feasible price interval. Credible news
that future retention will be below the minimum
 causes an immediate collapse through backward
unraveling. The model identifies a governance channel through which
insider unlocks can affect valuation without changing total token supply
or fundamental cash flows.
\end{abstract}

\begin{list}{}{}
    \item \textbf{Keywords}: rational bubbles; lockups; tokenomics; skin in the game.
    \item \textbf{JEL Classification}: G12, G32, D82, G34.
\end{list}
\newpage
\section{Introduction}
Who has an incentive to keep a market open when the asset traded in it
promises no cash flows? If continued resale depends on infrastructure
maintained by an issuer, foundation, or development team, the party that
bears the cost of upkeep need not capture its full benefit. Other holders
can sell into the market without paying for its preservation. Retained
ownership can therefore determine not only how an asset's value is
allocated, but whether a positive resale value can be sustained at all.

Crypto and platform tokens make this friction observable. They often pay
no dividend, so their price is a price of resale, and resale requires a
venue for listing, matching, and, in many cases, delegated market making.
Those activities are costly and are typically controlled by a foundation
or core team that also holds a locked allocation.
\cite{howellniessneryermack2020} study more than 1,500 initial coin
offerings. In their sample, 22~percent of issuers impose a vesting
schedule on insider tokens and 43~percent set tokens aside for a
foundation, bounty, or endowment. They interpret insider vesting,
following \cite{bravgompers2003} on IPO lock-ups, as a commitment device.
Some contracts make the lock-up part of the venue itself. Bancor reserved
funds for a market maker charged with maintaining price stability, funds
that could not be removed for a pre-specified period
\citep{howellniessneryermack2020}.

Throughout, we study a pure resale asset whose value depends on continued access to a functioning secondary market. We call the asset a \emph{token}, while abstracting from the transaction services, redemption rights, and profit claims that some actual tokens provide. 
The same mechanism applies to any store of value whose payoff is the
continuation of a resale institution.

The theory of zero-dividend assets takes that continuation as given. In
\cite{tirole1985asset} resale succeeds with certainty. In
\cite{weil1987confidence}, building on \cite{samuelson1958}, the market
remains open each period with an exogenous probability that Weil calls
confidence. Because nobody chooses whether the venue remains open, these
models cannot explain when confidence is incentive-compatible, nor
distinguish a market whose support is locked in from one whose support is
about to be released. The corporate-finance analysis of ownership and
lock-ups explains how retention disciplines a noncontractible action that
affects cash flows or listing quality
\citep{jensenmeckling1976,chemla2014skin,bravgompers2003}.
For a pure resale asset the value of the incentive instrument is itself
endogenous to the action it is meant to motivate. The sponsor maintains
the market because its retained tokens are valuable, and the tokens are
valuable because investors expect the sponsor to maintain the market.

We ask whether this feedback can support an asset that offers only the
prospect of resale, and what happens when the retained exposure is
released. The mechanism operates without a change in total token supply,
leverage, or fundamental cash flows. Instead, it turns on the exposure
retained by those responsible for maintaining the resale market.
IPO lock-ups offer a useful comparison: their expiration allows
insiders to sell. \cite{fieldhanka2001} examine 1,948 such agreements
and find a three-day abnormal return of $-1.5$~percent.
Unlike a pure resale asset, however, equity also represents a claim
on the firm's cash flows.

Token-unlock episodes illustrate how prices can respond before locked
holdings become transferable. In January 2023 the dYdX Foundation
announced that transfer restrictions on tokens issued to investors,
employees, and consultants would be extended, postponing the initial
release from 3~February 2023 to 1~December 2023.\footnote{dYdX Foundation,
``Lock-up extension for \$DYDX issued to investors, employees and
consultants,'' 24~January 2023, updated 31~January 2023,
\url{https://www.dydx.foundation/blog/lock-up-extension}.}
The token price rose when the delay became public. In late November
2023, DYDX, OP, and SUI fell ahead of large scheduled unlocks.
These episodes motivate a question beyond the price impact of
anticipated sales. If an unlock leads those responsible for maintaining
the resale market to reduce their exposure, it may also weaken their
incentive to keep that market operating. For a pure resale asset, the
risk is then not merely a lower price, but the loss of the institution
that makes resale possible.

We characterize when retained ownership supports a positive price, how
that restriction shapes live-path dynamics, and how an announced future
release unravels current demand.



Our framework is an overlapping-generations exchange economy. At each
active date, the young purchase a fixed supply of tokens from the old.
A predetermined fraction of the young belongs to a sponsor partnership.
Its members trade individually, but their holdings are subsequently
locked and managed collectively for the maintenance decision. After
trade, the sponsor chooses hidden maintenance that determines the
probability of another round of resale. The sponsor bears the entire
maintenance cost and internalizes only the benefit accruing to its
retained share. Investors rationally anticipate this
choice when buying the token. Survival is thus an incentive-compatible
outcome, rather than an exogenous confidence parameter.

The overlapping-generations structure isolates a recurring governance
problem that maintenance is controlled by a party whose
exposure remains locked until the survival uncertainty is resolved.
It does not require a new trading platform to be created each period.
A continuing foundation or core team motivates the institutional
interpretation, although the model does not describe a long-lived
organization's full financing or contracting problem. Relative to
\citet{tirole1985asset} and \citet{weil1987confidence}, the central change
is that someone must have an incentive to preserve the resale opportunity.

The first result establishes a minimum-retention threshold.
A positive stationary equilibrium  exists
if and only if retention exceeds this threshold. The threshold
compares the marginal cost of maintenance with the continuation value
consistent with household demand. Weil's minimum-confidence restriction
remains a necessary demand-side condition, but confidence must now also be
incentive compatible for the sponsor.

The second result concerns multiplicity and comparative statics. Under some convexity conditions, retention strictly above
the threshold supports exactly two positive stationary equilibria. The
lower equilibrium has a lower price and less maintenance, whereas the upper one
has a higher price and more maintenance. 

The third result characterizes equilibrium price dynamics conditional on
survival. Every positive initial price below the upper stationary price
converges to the lower stationary price. The upper stationary price is
itself sustainable; a higher initial price would eventually require a
resale payment that no future cohort can finance. The upper equilibrium
therefore determines the largest feasible initial price. Greater
retention expands the entire feasible price interval.


The fourth result concerns unlock announcements. Suppose investors learn
that, from a known finite date onward, the retained share will be
 below the minimum-retention threshold. No positive price can then be
supported after that date, whether stationary or not. The loss of future
resale unravels backward, and the bubble collapses when the news becomes
common knowledge. Smaller releases of retained exposure need not eliminate
all positive prices; they can instead reduce the set of sustainable
valuations. 

Our contribution is to connect a corporate-finance incentive constraint
to the existence and dynamics of a pure bubble. Token retention is not
new as a governance device. What is distinctive here is that the sponsor
protects no productive cash flow: its stake is valuable only if another
cohort is willing to buy. An unlock can therefore matter through the
exposure of the party responsible for continued trading, not merely
through additional selling pressure. Appendix~\ref{sec:policy} isolates
the resulting wedge between the sponsor's incentives and the joint
interests of current holders.

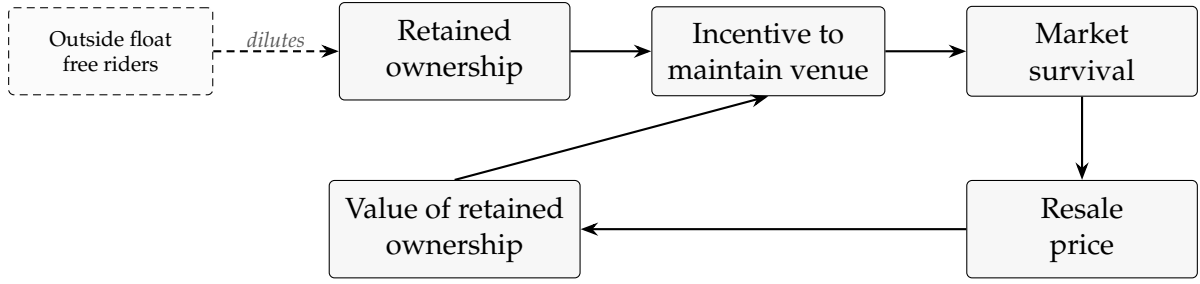
\begin{figure}[ht]
\centering
\begin{tikzpicture}[
    >=Stealth,
    box/.style={
        draw, rounded corners=2pt, align=center, inner sep=6pt, font=\small,
        minimum width=3.05cm, minimum height=1.15cm, fill=gray!6
    },
    side/.style={
        draw, rounded corners=2pt, align=center, inner sep=5pt, font=\scriptsize,
        minimum width=2.7cm, minimum height=1.15cm, fill=gray!3, densely dashed
    },
    arr/.style={->, thick},
    lab/.style={font=\scriptsize\itshape, text=black!65, inner sep=1pt}
]
\node[box] (retain) at (0,2.35) {Retained\\ownership};
\node[box] (maint)  at (4.15,2.35) {Incentive to\\maintain venue};
\node[box] (surv)   at (8.30,2.35) {Market\\survival};
\node[box] (value)  at (0,0) {Value of retained\\ownership};
\node[box] (price)  at (8.30,0) {Resale\\price};
\node[side] (float) at (-4.55,2.35) {Outside float\\free riders};

\draw[arr] (retain) -- (maint);
\draw[arr] (maint)  -- (surv);
\draw[arr] (surv)   -- (price);
\draw[arr] (price)  -- (value);
\draw[arr] (value.north) -- (maint.south);
\draw[arr, densely dashed] (float.east) -- node[lab, above] {dilutes} (retain.west);
\end{tikzpicture}
\caption{The maintenance and resale value.}
\begin{minipage}{1\textwidth}
{\vspace{.1in}
\footnotesize Note: The retained share is fixed. Solid arrows show how its value feeds back into maintenance through survival and resale prices. The dashed arrow is free-rider dilution from unlocked holdings. A larger outside float reduces the share of the maintenance benefit internalized by the sponsor.
}
\end{minipage}
\label{fig:mechanism_maint}
\end{figure}

Figure~\ref{fig:mechanism_maint} summarizes the feedback. Retained
ownership makes maintenance privately valuable; maintenance supports
survival; survival supports the resale price; and the price determines
the value of the sponsor's exposure. Passive holders benefit from
maintenance without bearing its cost.

Section~\ref{sec:literature} relates the paper to the literature.
Section~\ref{sec:environment} presents the model.
Section~\ref{sec:analysis} characterizes pricing, maintenance, stationary
equilibria, and dynamics, then discusses the token episodes.
Sections~\ref{sec:discussion} and~\ref{sec:conclusion} discuss the
interpretation and conclude. Appendix~\ref{app:effort_foundation} provides
an effort foundation, Appendix~\ref{sec:policy} studies the governance
externality, and Appendix~\ref{sec:general_uv} allows nonlinear old-age
utility. All proofs are collected in Appendix~\ref{app:proofs}.

\section{Related literature}
\label{sec:literature}

The paper connects ownership incentives and lockups in corporate finance
to the valuation of pure resale assets. The separation between control
and ownership creates an agency problem in \citet{jensenmeckling1976}.
\citet{chemla2014skin} study retention and moral hazard in securitization,
while \citet{maug1998,paganoroell1998} examine ownership and monitoring.
\citet{huvaras2025intermediary} analyze intermediary financing under
limited commitment. We use the same internalization logic, but the
sponsor's exposure is not a claim on a productive cash flow. Maintenance
preserves only an opportunity to resell.

The lockup literature provides a particularly useful comparison.
\citet{bravgompers2003} find support for a commitment explanation of IPO
lockups. \citet{fieldhanka2001} document trading-volume increases and
negative abnormal returns around lockup expirations.
\citet{ofekrichardson2003} connect the internet-stock boom and bust to
heterogeneous beliefs, short-sale restrictions, and the release of insider
shares. \citet{hongscheinkmanxiong2006} model how insider lockups and
anticipated changes in float affect speculative resale values under
disagreement and short-sale constraints. Here the relevant margin is the
controller's incentive to preserve trading, not the speculative resale
option generated by heterogeneous beliefs. This does not establish that illiquidity is always
desirable: informed blockholder trading can itself improve governance
\citep{edmans2009}.  Rather
than solve for an optimal lockup, we take the retention arrangement as given.

A second connection is to token valuation and platform finance.
\citet{congliwangs2021tokenomics,congliwangs2022platform} study token
adoption, valuation, and platform financing.
\citet{sockinxiong2023cryptocurrencies,sockinxiong2023tokenization}
examine cryptocurrency demand and decentralization through tokenization.
\citet{garrattwallace2018bitcoin} emphasize self-fulfilling valuations of
privately issued outside monies, and \citet{biais2023bitcoin} derive
cryptocurrency prices from net transactional benefits. Our benchmark
removes those benefits to isolate the incentive requirements for
continued resale.

Retained tokens already play an important role in financing and
governance. \citet{gantsoukalasnetessine2021ico,gantsoukalasnetessine2023dao}
analyze ICO design and incentives for platform building.
\citet{chodtrichakisyang2022} study financing, governance, and moral hazard
under tokenization. \citet{hanleeli2026} examine strategic token trading
and governance conflicts, including illiquidity and long-term incentives,
and \citet{canidio2026financial} connect developers' retained tokens to
investment incentives. These are close antecedents. Our contribution is
not the general idea that token holdings discipline behavior, but the
minimum-retention threshold, feasible price interval, and announcement
unraveling for an asset whose value comes entirely from resale.

Related platform-design questions include digital tokens and platform
building, corporate capture, tokens versus equity, staking, and token
distribution
\citep{limann2025digital,ferreira2023corporate,malinova2023tokenomics,congstaking2025,canidio2023token}.
Empirically, \citet{howellniessneryermack2020} study ICO financing and
subsequent real outcomes, highlighting the importance of disclosure and
credible project commitment. We do not model issuance proceeds or the
choice between token and equity financing. Retention is an exogenous
constraint on the sponsor's exposure during the maintenance decision.

The feedback between asset value and infrastructure incentives also
connects our analysis to blockchain security.
\citet{pagnotta2022decentralizing} jointly determine Bitcoin prices and
network security. \citet{budish2025trust} study the economic cost of
supporting blockchain trust and the incentives against attack.
\citet{biais2019blockchain,saleh2021blockchain} analyze consensus, including
the role of validators' ownership. These papers are important antecedents
to the idea that an asset's value helps sustain its infrastructure.
Our deviation is insufficient hidden upkeep rather than an attack or a
fork, and the token supplies neither transaction utility nor a
protocol-funded flow reward. We abstract from the contracting and mining
institutions studied by \citet{conghe2019blockchain,congheli2021pools}.
Evidence on crypto wash trading further cautions against using reported
volume alone to measure market activity \citep{congwash2023};
\citet{congbiaisetal2023advances} provide a broader perspective on this
research area.

Finally, the model builds on rational-bubble theory. The starting point
is the valuation of zero-dividend assets in overlapping generations
\citep{samuelson1958,tirole1985asset}. \citet{weil1987confidence} introduces
stochastic collapse and a minimum survival probability for stationary
existence. \citet{santoswoodford1997} provide a general asset-pricing
treatment, and \citet{martinventura2018guide} review the broader theory.
Debt and portfolio constraints affect bubble existence in
\citet{kocherlakota1992,kocherlakota2008,hugonnier2012}; liquidity,
investment, growth, and policy interact with bubbles in
\citet{farhitirole2012bubbly,martinventura2012growth,ikedaphan2019imbalances,asriyanfornaromartinventura2021monetary}.
\citet{endershakenes2021} emphasize the wealth available to absorb a
bubble. Household resources also bound resale payments here, but an
additional constraint comes from the incentives to maintain the market.

Other theories connect bubbles and crises to leverage and risk shifting
\citep{stiglitzweiss1981,allengale2000,barlevy2014leverage,ikedaphan2016toxic,benguiphan2018pledgeability,allenbarlevygale2022},
synchronization \citep{abreubrunnermeier2003}, or the dynamics of asset
quality \citep{fukui2018asset,caramp2025sowing};
\citet{goldstein2025fragility} review related sources of financial
fragility. Our channel requires neither debt nor a deterioration in the
token's fundamental payoff. It is the incentive-compatible probability
of continued resale that changes. The lockup commits exposure, not the
hidden action or future prices.

\section{The model}
\label{sec:environment}

\subsection{Demographics and preferences}

Time is discrete and runs forever. At each date $t$, a unit mass of agents
is born and lives for two periods. Each generation is endowed with $e_1>0$ units of perishable consumption
good when young and $e_2\geq 0$ units when old. Preferences are
\[
    u(c^y_{it})
    +\gamma\, E_t[c^o_{i,t+1}]
    \;-\;
    \text{maintenance disutility, if any},
\]
where any maintenance disutility is borne by the sponsor partnership introduced below.

\begin{assumption}[Young-age utility]
\label{ass:linear_v_utility}
The function $u:(0,e_1]\to\mathbb{R}$ is twice continuously
differentiable and satisfies
\[
    u'(\cdot)>0,
    \qquad
    u''(\cdot)<0,
    \qquad
    \lim_{c\downarrow 0}u'(c)=+\infty.
\]
Moreover,
\begin{equation}
    u'(e_1)<\gamma.
    \label{eq:linear_v_dynamic_inefficiency}
\end{equation}
\end{assumption}

Condition~\eqref{eq:linear_v_dynamic_inefficiency} is the baseline
dynamic-inefficiency restriction. At the endowment allocation, the
marginal utility cost of transferring one unit out of young-age
consumption is smaller than its old-age utility benefit. It therefore
leaves room for a risk-free store of value. 
The condition plays the role of the interest-rate-versus-growth-rate
restriction in classical overlapping-generations bubble models
\citep{samuelson1958,tirole1985asset}.

Old-age utility is linear, and $\gamma>0$ measures the weight on old-age
consumption. In Appendix~\ref{sec:general_uv}, we allow
the old-age utility
to be an arbitrary increasing concave function. All our results can go through under some additional technical assumptions.

\subsection{The token}

There is a fixed unit supply of a divisible token. The token pays no
dividend, provides no transaction service or collateral value, carries no
redemption right, and conveys no claim on profits. Any positive price
therefore reflects resale value alone, as in the pure-bubble environments
of \citet{tirole1985asset} and \citet{weil1987confidence}.

Let $Z_t\in\{0,1\}$ indicate whether the resale venue is active.
Collapse is absorbing: if $Z_t=0$, then $Z_{t+j}=0$ for all $j\geq1$.
Conditional on $Z_t=1$, let $B_t>0$ be the price at which the young buy
the token from the old. A holder who buys at date $t$ receives
$B_{t+1}$ upon resale next period if the venue remains active, and zero
otherwise. Sustaining
the bubble creates no aggregate resources, and it organizes a transfer from
the young to the old.

\subsection{The sponsor partnership and maintenance technology}

A predetermined share $ s \in(0,1]$ of each young cohort is designated
as the sponsor partnership. The remaining share $1- s $ consists of
passive investors. Designation is imposed by the institution operating the
market rather than chosen by individual agents. All agents trade
individually and anonymously. In the symmetric allocation derived below,
each holds one token per capita, so $ s $ is also the partnership's
share of aggregate token ownership.

After trade closes, the sponsor's holdings are locked and its manager
chooses the survival probability\footnote{Appendix~\ref{app:effort_foundation}
derives this reduced-form choice from hidden operational effort.}
\[
q_t\equiv\Pr(Z_{t+1}=1\mid Z_t=1)\in[0,1).
\]
The manager acts on behalf of the partnership. Its positions cannot be
sold, hedged, or pledged before the survival outcome is realized. This
restriction preserves the exposure that motivates maintenance. Throughout
the baseline, retention is constant and publicly known. 

\begin{assumption}[Maintenance technology]
\label{ass:maintenance}

The maintenance cost function
\[
\psi:[0,1)\rightarrow\mathbb{R}_+
\]
is twice continuously differentiable and satisfies
\[
    \psi(0)=0,
    \qquad
    \psi'(0)\geq 0,
    \qquad
    \psi''(\cdot)>0,
    \qquad
    \lim_{q\rightarrow 1}\psi'(q)=+\infty.
\]
\end{assumption}

The function $\psi$ is the cost of achieving survival probability $q$.
Strict convexity captures increasing marginal difficulty in improving
survival, and the divergence of $\psi'$ at one rules out perfect survival
at a finite marginal benefit. The condition $\psi'(0)\geq0$ allows a
positive marginal cost at zero maintenance.

The total maintenance cost does not depend on the sponsor's ownership
share. It represents an infrastructure burden that must be borne regardless
of how ownership is divided. Consequently, the sponsor pays the full
cost but captures only the fraction $ s $ of holders' continuation
benefit. Maintenance disutility is shared within the partnership through
an allocation that does not depend on any individual member's token
holdings. This assumption separates atomistic portfolio choices from the
subsequent collective maintenance decision.

Maintenance is a utility cost rather than a goods cost. Aggregate goods
feasibility is therefore
\begin{equation}
\label{eq:feasibility}
\int_0^1 c^y_{it}\,di+\int_0^1 c^o_{it}\,di=e_1+e_2.
\end{equation}
A goods-cost specification would additionally link maintenance to
available consumption. We abstract from that resource channel to isolate
the effect of delegated control and incomplete ownership retention.

\Needspace{12\baselineskip}
\subsection{Timing}

Conditional on $Z_t=1$, events within date $t$ follow the sequence in
Figure~\ref{fig:timing_maint}.
\begin{enumerate}
    \item The old hold the token, and a competitive resale price $B_t$
    is determined in equilibrium.
    \item The market clears: the young acquire the unit supply at
    cost $B_t$, and the old receive $B_t$ and exit.
    \item Sponsor positions are locked. The sponsor chooses the hidden
    survival probability $q_t$.
    \item Nature draws survival: with probability $q_t$ the market
    remains live into $t+1$; otherwise it collapses forever.
\end{enumerate}

When maintenance is chosen, the current price has been paid and
portfolios are fixed. The price $B_t$ reflects anticipated upkeep but
cannot be made contingent on the sponsor's subsequent hidden action.
The lock-up keeps the sponsor exposed to the next resale payoff during
this decision. The old have already sold and exited, so maintenance is
chosen for the benefit of current, not former, holders.


\begin{figure}[ht]
\centering
\begin{tikzpicture}[
    >=Stealth,
    stage/.style={draw, rounded corners=2pt, align=center,
                  font=\footnotesize, inner sep=5pt,
                  text width=2.65cm, minimum height=1.3cm,
                  fill=gray!5},
    arr/.style={->, thick},
    note/.style={font=\scriptsize\itshape, text=black!60, align=center}
]
\node[stage] (p1) {1.\\Price forms};
\node[stage, right=0.45cm of p1] (p2)
    {2.\\Market clears};
\node[stage, right=0.45cm of p2] (p3)
    {3.\\Positions locked\\[1pt]{\scriptsize Sponsor chooses $q_t$}};
\node[stage, right=0.45cm of p3] (p4)
    {4.\\Survival draw\\[1pt]{\scriptsize Resale or collapse}};
\draw[arr] (p1) -- (p2);
\draw[arr] (p2) -- (p3);
\draw[arr] (p3) -- (p4);
\node[note] at ($(p2.south)!0.5!(p3.south)+(0,-0.55)$)
    {Hidden maintenance follows trade; the survival draw comes last};
\end{tikzpicture}
\caption{Within-period timing on a live date. }
\begin{minipage}{1\textwidth}
{\vspace{.1in}
\footnotesize Note: Trading and clearing
precede the hidden maintenance choice, and survival is realized afterward.
}
\end{minipage}
\label{fig:timing_maint}
\end{figure}

We restrict attention to rational-expectations equilibria with
deterministic price sequences conditional on survival. At each live date,
households correctly anticipate the sponsor's choice $q_t$ and the
next-period price $B_{t+1}$ if the venue survives. The realized resale
payment is $B_{t+1}$ with probability $q_t$ and zero otherwise.
Prices and maintenance can vary over time, but there is no additional
aggregate uncertainty or extrinsic randomization. Once the venue
collapses, the token is permanently worthless.

\section{Equilibrium analysis}
\label{sec:analysis}

\subsection{Household pricing}
\label{sec:households}
A young agent who buys $a_{it}$ units of the token at price $B_t>0$
consumes $c_{it}^y=e_1-B_ta_{it}$ when young. When old, the agent consumes
\[
    c_{i,t+1}^o=
    \begin{cases}
        e_2+B_{t+1}a_{it}, & Z_{t+1}=1,\\
        e_2, & Z_{t+1}=0.
    \end{cases}
\]
Expected old-age consumption is therefore
$e_2+q_tB_{t+1}a_{it}$. Maintenance disutility, if any, is additive
and independent of an individual agent's holdings, so it does not enter
the portfolio margin. Taking $q_t$ and the live continuation price
$B_{t+1}$ as given, each young agent solves
\[
    \max_{0\leq a_{it}<e_1/B_t}
    \left\{u(e_1-B_ta_{it})
    +\gamma\bigl(e_2+q_tB_{t+1}a_{it}\bigr)\right\}.
\]
The objective is strictly concave. At an interior optimum,
\begin{equation}
\label{eq:household_foc}
    B_tu'(e_1-B_ta_{it})=\gamma q_tB_{t+1}.
\end{equation}
The left-hand side is the marginal utility cost of an additional token, and
the right-hand side is its expected old-age benefit. The corner solution, i.e., the bubbleless
equilibrium is treated separately below.

All young agents solve the same strictly concave portfolio problem.
Token-market clearing, $\int_0^1a_{it}\,di=1$, therefore implies
$a_{it}=1$ for every agent in a symmetric positive-price equilibrium.
The sponsor partnership consequently owns the share $ s $ of the
unit supply. Substituting into the first-order condition gives
\begin{equation}
    B_tu'(e_1-B_t)=\gamma q_tB_{t+1}.
    \label{eq:linear_v_nonstationary_pricing}
\end{equation}

In a stationary live state $B_t=B_{t+1}=B>0$, so
\eqref{eq:linear_v_nonstationary_pricing} reduces to
\begin{equation}
    u'(e_1-B)=\gamma q.
    \label{eq:linear_v_stationary_pricing_condition}
\end{equation}
Define the stationary survival threshold and the limiting stationary price
\begin{equation}
    \underline q
    \equiv
    \frac{u'(e_1)}{\gamma}
    \in(0,1),
    \qquad
    \overline B
    \equiv
    e_1-(u')^{-1}(\gamma)
    \in(0,e_1).
    \label{eq:linear_v_bounds}
\end{equation}
The stationary bubble price is given by
\begin{equation}
    \mathcal{B}(q)
    \equiv
    e_1-(u')^{-1}(\gamma q),
    \qquad
    q\in(\underline q,1).
    \label{eq:linear_v_pricing_locus}
\end{equation}

\begin{lemma}[Generalized Weil restriction]
\label{lem:linear_v_weil}
Every positive stationary
live path satisfies $q>\underline q$ and $B\in(0,\overline B)$.
The stationary bubble price $\mathcal{B}$ is continuous and strictly increasing.
\end{lemma}
The proofs for the baseline model are collected in Appendix~\ref{app:proofs_baseline}.

The lemma is Weil's restriction in general form. The threshold
$\underline q=u'(e_1)/\gamma$ is the infimum of the survival probabilities
consistent with a positive stationary price. At or below this threshold,
no positive stationary price clears the token market. This restriction is
stationary, since away from a steady state the household demand depends on both
$B_t$ and $B_{t+1}$. The monotonicity of $\mathcal{B}$ states that a
higher stationary survival probability supports a higher stationary price.

Sponsor members satisfy the same portfolio first-order condition.
Each member is atomistic and cannot individually affect the aggregate
survival probability or the partnership's maintenance cost. The member
therefore takes $q_t$ as given at the trading stage, while the manager
chooses maintenance collectively after trade. Rational expectations
require the anticipated probability to equal that subsequent choice.

\subsection{The sponsor's post-trade problem}
\label{sec:sponsor}

Once trade has closed, the price paid by the young is sunk and portfolio
positions are fixed. At date $t$, the sponsor manager chooses the survival
probability protecting the members' resale payoff at date $t+1$. Aggregating
over the sponsor partnership, the manager solves

\begin{equation}
\label{eq:sponsor_problem}
    \max_{q_t\in[0,1)}
    \left\{
     s \gamma q_tB_{t+1}-\psi(q_t)
    \right\}.
\end{equation}
The objective is strictly concave because $\psi''(\cdot)>0$. The first-order
condition for an interior solution is
\begin{equation}
\label{eq:sponsor_foc}
    \psi'(q_t)= s \gamma B_{t+1} .
\end{equation}
Allowing for a possible corner solution at zero maintenance, the sponsor
best response is therefore
\begin{equation}
\label{eq:qD_general}
\mathbf{q}(B_{t+1})
\equiv
\begin{cases}
0,
&
\text{if }  s \gamma B_{t+1}\leq \psi'(0),
\\[6pt]
(\psi')^{-1}( s \gamma B_{t+1}),
&
\text{if }  s \gamma B_{t+1}>\psi'(0).
\end{cases}
\end{equation}

The problem also admits a dual representation. Let $\psi^*$ denote the
Fenchel conjugate of $\psi$:
\[
    \psi^*(x)
    \equiv
    \sup_{q\in[0,1)}
    \{xq-\psi(q)\}.
\]
The sponsor's optimized value from maintenance is therefore

\[
V(B_{t+1})=\psi^*( s \gamma B_{t+1}).
\]
The sponsor's marginal value of survival is $s\gamma B_{t+1}$, whereas
all current holders together value it at $\gamma B_{t+1}$. Passive
holders benefit from upkeep, but their gains do not enter the sponsor's
objective. This creates a wedge between the valuation of the  partnership versus all token holders.  In appendix \ref{sec:policy}, we analyze this governance externality.

Holding the continuation price fixed, a lower retained stake weakens
maintenance incentives without changing the cost of upkeep. Maintenance
falls weakly, and strictly when the optimum is interior. This is the
internalization force familiar from retention and monitoring models
\citep{chemla2014skin}. 
At the limiting case
$ s =0$, maintenance is zero at every continuation price, so a positive
bubble cannot be sustained.

\subsection{Stationary equilibrium}
\label{sec:equilibrium}

\begin{definition}[Live-path stationary equilibrium]
\label{def:equilibrium}
A \emph{live-path stationary equilibrium} is a pair $(B,q)$ with stationary bubble price $B>0$ and stationary survival probability $q\in(0,1)$ such that
\begin{enumerate}
    \item \emph{Household optimality:}
    $B=\mathcal{B}(q)$;
    \item \emph{Sponsor optimality:} $q=\mathbf{q}(B)$;
    \item \emph{Market clearing:} each young agent holds one token per capita, and goods feasibility~\eqref{eq:feasibility} holds.
\end{enumerate}
\end{definition}

A \emph{bubbleless equilibrium} has $B=q=0$, $c^y=e_1$, and
$c^o=e_2$ at every date. It always exists that a zero continuation price
induces zero maintenance, and without a positive expected resale payoff
no household pays a positive price. The question is whether the economy
also admits a positive stationary equilibrium.

\begin{theorem}[Existence and minimum retention]
\label{thm:linear_v_existence}
Under Assumptions~\ref{ass:linear_v_utility} and~\ref{ass:maintenance},
a positive live-path stationary equilibrium exists if and only if
\[
     s \geq s _{\min}
    \equiv
    \min_{q\in(\underline q,1)}
    \frac{\psi'(q)}{\gamma [e_1-(u')^{-1}(\gamma q)]}.
\]
Some feasible retention level $ s \in(0,1]$ can support a positive
stationary price if and only if $ s _{\min}\leq1$.
\end{theorem}

The threshold $ s _{\min}$ is the smallest stake that makes the
sponsor willing to supply a survival probability supported by household
demand. The numerator $\psi'(q)$ is marginal maintenance cost, and the
denominator $\gamma\mathcal{B}(q)$ is the marginal value of survival
to \emph{all} holders at the associated stationary price. Their ratio
is the ownership share needed to align the sponsor's private marginal
benefit with that cost. Minimizing over $q$ identifies the least retention
consistent with a positive stationary bubble.

The following curvature restriction gives a sharper characterization.
\begin{assumption}[Curvature]
\label{ass:linear_v_curvature}
The marginal utility function $u'$ and the marginal maintenance cost
function $\psi'$ are weakly convex.
\end{assumption}

Marginal utility being convex is satisfied by many
standard utility functions (logarithmic, CRRA, CARA). Weak convexity of $\psi'$ is the
usual regularity condition on the cost of raising a success
probability in moral-hazard models with a continuum of effort
\citep{laffont2002theory}.\footnote{In \cite[pp.~190--191]{laffont2002theory}, the agent chooses effort
$e\in[0,1]$ at cost $\psi(e)$ with $\psi',\psi''>0$ and
$\lim_{e\to1}\psi'(e)=+\infty$; $\psi'''\geq0$ is imposed so that
the information rent $e\psi'(e)-\psi(e)$ is convex and the
principal's problem is concave. Our $q$ is similar to the setup of \cite{laffont2002theory}.}
It holds for $\psi(q)\propto -\log(1-q)$, $\psi(q)\propto q/(1-q)$, and
$\psi(q)\propto (1-q)^{-\eta}$ with $\eta>0$.
These specifications also satisfy $\psi(0)=0$ and
$\lim_{q\to 1}\psi'(q)=+\infty$.

\begin{proposition}[Multiplicity]
\label{prop:linear_v_multiplicity}
Suppose Assumptions~\ref{ass:linear_v_utility}, \ref{ass:maintenance}, and~\ref{ass:linear_v_curvature} hold. If $ s > s _{\min}$, there exist exactly two positive
stationary equilibria.
\end{proposition}
Write the two equilibria as $(B_-,q_-)$ and $(B_+,q_+)$, ordered so
that $B_-<B_+$ and $q_-<q_+$. In either equilibrium, beliefs about
resale and maintenance incentives are mutually consistent. On the lower
branch, a lower anticipated price gives the sponsor a weaker maintenance
incentive, and the resulting lower survival probability supports that
price. On the upper branch, a higher price and stronger maintenance
reinforce one another. Both survival probabilities nevertheless exceed
the household threshold $\underline q$, and the lower branch is not a
zero-maintenance outcome.


\subsection{Dynamics of the bubble price}
\label{sec:linear_v_forward_dynamics}

We now study dynamic price paths conditional on survival. Household
optimality and market clearing give
\eqref{eq:linear_v_nonstationary_pricing}. Along a positive live path,
$q_t>0$, so sponsor optimality is characterized by the interior
condition~\eqref{eq:sponsor_foc}. Eliminating $B_{t+1}$ gives
\begin{equation}
B_tu'(e_1-B_t)=\frac{q_t\psi'(q_t)}{ s }.
\label{eq:linear_v_current_price_maintenance}
\end{equation}
The function $q\mapsto q\psi'(q)$ is continuous and strictly increasing
from zero to infinity. Thus each current price $B_t\in(0,e_1)$ pins
down a unique positive maintenance choice and an associated continuation
price,
\[
B_{t+1}=T(B_t)\equiv\frac{\psi'(q_t)}{ s \gamma},
\qquad
q_t\text{ solves }\eqref{eq:linear_v_current_price_maintenance}.
\]
The map $T$ is an algebraic candidate transition on $(0,e_1)$.
A candidate continuation is \textit{feasible} only if $T(B_t)<e_1$.
The map is continuous and strictly increasing. When
$ s > s _{\min}$, its only positive fixed points are $B_-$ and
$B_+$.

\begin{theorem}[Dynamics of the bubble price]\label{thm:linear_v_forward_dynamics}
Suppose Assumptions~\ref{ass:linear_v_utility}, \ref{ass:maintenance}, and~\ref{ass:linear_v_curvature} hold and $ s > s _{\min}$.
Conditional on the market remaining live, if
$B_0\in(0,B_+)$, $B_t\to B_-$. If
$B_0\in(B_+,e_1)$, the resulting path is infeasible. 
\end{theorem}

The intuition is as follows.
The household pricing condition can be written as
\[
\frac{B_{t+1}}{B_t}=\frac{u'(e_1-B_t)}{\gamma q_t}.
\]
A higher current price absorbs more of the young cohort's endowment and
raises the marginal utility cost of saving through the token. Supporting
that price therefore requires a larger expected resale payoff. The sponsor
can improve the probability of resale, but it does so only in response to
the value of its locked stake, and at an increasing marginal cost. The
price transition reconciles these two requirements, the payoff households
need and the maintenance the sponsor is willing to supply.

At a stationary price $B_i$, differentiating household pricing after
substituting the sponsor's best response gives
\[
T'(B_i)=
\frac{u'(e_1-B_i)-B_i u''(e_1-B_i)}
{\gamma[\mathbf{q}(B_i)+B_i\mathbf{q}'(B_i)]},
\qquad i\in\{-,+\}.
\]
The numerator measures how a higher current price raises the marginal
utility cost of an additional token, evaluated at market-clearing holdings. The denominator is the increase
in its expected resale payoff when the continuation price rises, including
the induced maintenance response. At the lower equilibrium the latter
response is stronger, so a price deviation requires a smaller continuation
deviation and contracts over time. At the upper equilibrium the ranking is
reversed, so deviations expand. Thus, the lower equilibrium is stable while the upper equilibrium is unstable.

The global movements follow the same logic. Below $B_-$, the implied
resale opportunity is too weak to support a constant price, so households
require appreciation conditional on survival. Between $B_-$ and $B_+$,
maintenance supports sufficient expected resale value for the price to
decline toward $B_-$. Above $B_+$, continued appreciation would be
required, but it cannot remain feasible. The next generation has only
$e_1$ goods, and its marginal utility of the consumption left after buying
the token diverges as the price approaches that bound.

Note that convergence along a live path does not imply permanent
survival, because the venue can fail at any date through the survival draw. Moreover, the movements are equilibrium restrictions. In particular, an infeasible candidate
sequence above $B_+$ is not an equilibrium boom that investors knowingly
ride until an inevitable terminal crash. Rational expectations rule out
its initial price. The theorem thus identifies the exact set of feasible prices. 

\Needspace{10\baselineskip}
\begin{corollary}[Retention and the feasible price range]
\label{cor:viable_range}
Suppose Assumptions~\ref{ass:linear_v_utility}, \ref{ass:maintenance},
and~\ref{ass:linear_v_curvature} hold. For each
$ s > s _{\min}$, the feasible price range is
\[
\mathcal{V}( s )=(0,B_+( s )].
\]
If $ s _{\min}< s '< s \leq1$, then
$\mathcal{V}( s ')\subsetneq\mathcal{V}( s )$.
\end{corollary}
In practice, an unlock or vesting releases the maintenance provider's
previously committed token exposure, leading to a reduction in
retention $s$. Corollary \ref{cor:viable_range} shows that a smaller retained stake tightens the price bound.
At a fixed continuation price, a smaller stake lowers the sponsor's
private return to maintenance without reducing its cost. The same force
can be viewed from a fixed \emph{current} price. Equation~
\eqref{eq:linear_v_current_price_maintenance} then implies lower
maintenance when retention falls. Household pricing consequently requires
a higher continuation price to compensate for the lower survival
probability. A reduction in retention thus raises the future resale price
needed to justify a given current payment, even as it lowers the largest
price that can be sustained indefinitely. Prices near the old upper bound
can cease to be feasible.


A stronger conclusion follows when retention becomes permanently smaller
than $ s _{\min}$. When that happens, positive stationary equilibria cease to exist and the increasing transition map has no positive
fixed point. Every positive candidate price then requires
further appreciation, and its continuation reaches the endowment bound
after finitely many steps. Denote that date by $T$.

Forward-looking households cannot justify a current payment by a resale
price that no future cohort can pay in equilibrium. If a zero resale value
at date $T$ is common knowledge, a young buyer at $T-1$ receives no
payoff from holding the token and therefore pays nothing. The same logic
applies at $T-2$ and recursively at every earlier date at which this news
is known. Because the token has no dividend or other direct benefit,
nothing remains to support its current valuation. The following result
makes this announcement effect precise.

\Needspace{14\baselineskip}
\begin{theorem}[Unlock announcements and immediate bubble collapse]
\label{thm:retention_crash}
Suppose Assumptions~\ref{ass:linear_v_utility} and~\ref{ass:maintenance}
hold. At a live date $\tau$, before trade, it becomes common knowledge
that there are a finite date $t_*\geq\tau$ and a constant
$\widehat s\in(0,1]$ with $\widehat s<s_{\min}$ such that the retained
share satisfies
\[
    0\leq s_n\leq\widehat s
    \qquad\text{for every }n\geq t_*.
\]
Then $B_\tau=0$. Any positive valuation prevailing before the
announcement collapses when the news arrives.
\end{theorem}

The result follows from the resale requirement, not from an assumption
that maintenance immediately stops. Even below $s_{\min}$, a sufficiently
high hypothetical continuation price can induce positive maintenance.
But no sequence of such prices can satisfy both the sponsor's incentive
constraint and investors' budget and optimality conditions indefinitely.
Hence, conditional on reaching $t_*$, the token has no positive
equilibrium resale value. A buyer at $t_*-1$ then has nothing to resell,
so that cohort also pays zero. The same reasoning applies at every
earlier date at which the future regime is known, including $\tau$.

This is an announcement effect, not a prediction of a fully anticipated
crash on the mechanical unlock date. A new vesting agreement, a revised
release schedule, or credible news that previously committed exposure
will be withdrawn can reveal the loss of support before any tokens are
sold. Conversely, if a schedule satisfying the theorem is already common
knowledge when trading begins, a positive bubble cannot exist in the
first place. The bound $s_n\leq\widehat s<s_{\min}$ includes a one-time
permanent reduction and subsequent rounds of further unlocking. A
smaller release that leaves sufficient retention does not by itself
imply a collapse to zero. Corollary~\ref{cor:viable_range} instead
describes the contraction of the feasible price range.

Announcement-driven collapse differs from eventual stochastic failure.
At a stationary live equilibrium, the probability of surviving another
$n$ dates is $q^n$, which tends to zero because $q<1$. Collapse thus
occurs almost surely in finite time even though the token has a positive
price after every surviving history
\citep{weil1987confidence,blanchardwatson1982}.
The same distinction applies to the nonstationary paths in
Theorem~\ref{thm:linear_v_forward_dynamics}: prices are bounded above by
$B_+$ and one-period survival probabilities by $q_+<1$. Almost-sure
eventual failure does not supply a known last trading date. In contrast,
Theorem~\ref{thm:retention_crash} supplies a known future date at which
every equilibrium resale value is zero. Conditional
price convergence, stochastic failure, and announcement-driven
unraveling are therefore different equilibrium statements.

\subsection{Token unlocks and the exposure of insiders}
\label{sec:unlock_cases}

The preceding results link insider retention to the sustainability of
resale values. Corollary~\ref{cor:viable_range} shows that a lower
retained share reduces the maximum sustainable price when retention
remains above $s_{\min}$. Theorem~\ref{thm:retention_crash} gives a
sharper prediction: credible news of a permanent future reduction
below the threshold eliminates positive resale values immediately.
The mechanism operates through maintenance incentives, without
requiring a change in total token supply or fundamental cash flows.
The following episodes illustrate price movements ahead of unlocks
and in response to changes in lock-up arrangements.

Prices can decline before previously locked tokens become
transferable. On 21 October 2022, AXS was reported to have fallen
more than 12.5 percent over the preceding 24 hours ahead of a
late-October unlock.\footnote{Adam James, Axie Infinity Shards
price drops sharply ahead of early investor token unlock,
\textit{The Block}, 21 October 2022,
\url{https://www.theblock.co/news/web3/2022-10-21-axie-infinity-shards-price-drops-sharply-ahead-of-early-investor-token-unlock-178860}.}
On 27 November 2023, DYDX, OP, and SUI fell ahead of unlocks
scheduled to make almost \$650 million of previously locked tokens
transferable that week. DYDX fell 7 percent ahead of a 1 December
release of about \$480 million, reported as nearly doubling
circulating supply. OP fell nearly 6 percent ahead of a
30 November unlock of about \$40 million, while SUI fell
8.8 percent with about \$48 million scheduled for release during
the week. All three declines exceeded the concurrent 1.5 percent
decline in the CoinDesk Market Index.\footnote{CoinDesk,
Altcoin rally cools as massive \$650M worth of token unlocks
loom over crypto market, 27 November 2023,
\url{https://www.coindesk.com/markets/2023/11/27/altcoin-rally-cools-as-massive-650m-worth-of-token-unlocks-loom-over-crypto-market}.}

An extension of transfer restrictions was accompanied by a price
increase. In January 2023, the dYdX Foundation announced an
extension of lock-ups on tokens issued to investors, employees,
and consultants. The change postponed an initial release of
150 million tokens from 3 February 2023 to 1 December 2023,
with subsequent monthly releases continuing through
June 2026.\footnote{dYdX Foundation, Lock-up extension for
\$DYDX issued to investors, employees and consultants,
24 January 2023, updated 31 January 2023,
\url{https://www.dydx.foundation/blog/lock-up-extension}.}
On 25 January, DYDX rose as much as 21 percent as news of the
delay became public.\footnote{Oliver Knight, DYDX token surges
as unlock gets delayed until December, \textit{CoinDesk},
25 January 2023,
\url{https://www.coindesk.com/business/2023/01/25/dydx-token-surges-21-as-report-says-token-unlock-delayed-until-december}.}
The price response thus preceded the originally scheduled release.

Admittedly, these patterns are not unique to the maintenance channel.
An anticipated expansion in the tradable float can increase
expected selling pressure. Investors who expect weaker future
resale prices may sell in advance, putting downward pressure on
the current price before the unlocked tokens reach the market.
Conversely, postponing an unlock can raise the price by deferring
expected sales. These responses can arise without any change in
maintenance incentives, even when total token supply is fixed.
Our model isolates a different channel: an unlock that reduces
the maintainer's effective exposure weakens its incentive to
preserve future resale. Transferability alone does not establish
such a reduction, since insiders may continue to hold the tokens.
The episodes therefore motivate the mechanism rather than
identify it. Distinguishing the channels would require evidence
on changes in the maintainer's unhedged exposure and maintenance
activity, alongside expected sales and market liquidity.



\section{Discussion}
\label{sec:discussion}

The benchmark isolates a governance component of resale value.
Maintenance changes the probability of another trading opportunity, not
the token's dividend or direct service payoff. Introducing redemption,
transaction utility, or distributions from a productive platform would
add a fundamental component. In that setting, a loss of maintenance
incentives need not drive the entire token price to zero.

The relevant empirical state variable is the maintenance provider's
effective, unhedged exposure, not the aggregate fraction of tokens
reported as locked. An unlock for passive financial investors primarily
changes their trading opportunities; an unlock that releases the
controller's exposure can additionally change maintenance incentives.
This distinction suggests comparing unlocks by recipient identity,
residual insider holdings, alternative sources of operating finance,
and subsequent measures of development or market support. The model
predicts a tighter feasible price bound when committed exposure falls.

Retention is a commitment to exposure, not to a particular action or to
successful survival. The baseline excludes separate compensation for
maintenance, external subsidies, and franchise or reputational rents.
These could sustain upkeep after token exposure declines, but their
funding and enforcement would have to be modeled explicitly. Similarly,
a persistent sponsor would introduce intertemporal incentives beyond the
recurring post-trade problem studied here.

Longer lockups are therefore not an unconditional policy prescription.
They preserve the exposure that motivates maintenance in this model,
but can have costs or benefits through channels that the model omits.
For example, liquidity can facilitate monitoring and informed
blockholder trading \citep{maug1998,edmans2009}. A theory of the optimal
vesting schedule would need to trade off these forces and allow the
sponsor's financing and trading decisions to respond. Our comparative
statics hold the maintenance technology and the available incentive
instruments fixed.

\section{Conclusion}
\label{sec:conclusion}

A pure bubble can depend on an institution that is costly to maintain.
When the party responsible for that institution retains only part of the
asset, its maintenance choice internalizes only part of the benefit from
continued resale. Survival is then an equilibrium outcome of a governance
problem rather than an exogenous confidence parameter.

The interaction between household demand and maintenance incentives
produces a minimum-retention threshold for stationary existence. Under
additional curvature restrictions, retention above that threshold supports
two positive stationary equilibria with opposite comparative statics.
Dynamic analysis distinguishes their roles. The lower price attracts all
positive live paths below the upper price, while the upper price is the
maximum feasible initial price. A larger retained stake expands this
feasible interval.

The loss of governance support can also have an immediate valuation effect.
A permanent reduction in retention below the threshold eliminates all
positive continuation values and causes immediate backward unraveling.
The model thus identifies a governance channel through which an unlock
can affect valuation even without new issuance or a change in
fundamental cash flows. The relevant question is not only how many tokens
become tradable, but whether the party responsible for continued resale
retains enough exposure to support it.

\newpage
\appendix
\numberwithin{equation}{section}
\section{An effort foundation for the survival choice}
\label{app:effort_foundation}

This appendix provides an effort-based foundation for the reduced-form
maintenance technology in the main text. Maintenance is interpreted as
hidden operational effort that raises the probability of continued resale. We show how
an increasing, weakly concave survival technology and a strictly convex
effort cost generate a strictly convex cost of achieving a given survival
probability. Under the conditions stated below, this induced cost
satisfies Assumption~\ref{ass:maintenance}. The direct choice of survival
can therefore be understood as a reparameterization of an underlying
effort problem, preserving the sponsor's maintenance incentives.

Let the sponsor choose
$e\geq0$ at cost $C(e)$, and let survival be $q=\varphi(e)$, where
$\varphi(0)=0$, $\varphi'(e)>0$, $\varphi''(e)\leq0$, and $\varphi(e)\to1$ as
$e\to\infty$. Suppose also that $C(0)=0$, $C'(e)\geq0$, and
$C''(e)>0$. Because $\varphi$ is strictly increasing, define
\[
 \psi(q)\equiv C\bigl(\varphi^{-1}(q)\bigr).
\]
Choosing effort to maximize
$ s \gamma \varphi(e)B-C(e)$ is then equivalent to choosing survival to
maximize the sponsor objective in \eqref{eq:sponsor_problem}. Writing
$e=\varphi^{-1}(q)$, the derivatives of the reduced cost are
\begin{align*}
\psi'(q)&=\frac{C'(e)}{\varphi'(e)},
\\
 \psi''(q)&=
 \frac{C''(e)\varphi'(e)-C'(e)\varphi''(e)}{[\varphi'(e)]^3}>0.
\end{align*}
The final inequality follows from strictly convex effort cost and weakly
concave survival production. The marginal-cost boundary condition in
Assumption~\ref{ass:maintenance} follows whenever
$C'(e)/\varphi'(e)\to\infty$ as $e\to\infty$. If $C'(0)=0$, then $\psi'(0)=0$; if $C'(0)>0$, then
$\psi'(0)>0$. Both cases are permitted by Assumption~\ref{ass:maintenance}. Thus choosing $q$ directly is a reduced-form representation of maintenance
effort, not an assumption that the sponsor literally purchases a
probability.

\section{The governance externality}
\label{sec:policy}
This appendix examines the governance externality created by incomplete
ownership retention. The sponsor bears the full cost of maintenance,
while part of the resulting benefit accrues to passive token holders.
Holding the continuation price and post-trade portfolios fixed, we
compare the sponsor's choice with the maintenance decision that maximizes
the joint continuation value of all current holders. This comparison
shows why delegated control can lead to insufficient maintenance and
how greater retention narrows the incentive wedge. 

Fix a continuation price $B$ and the post-trade portfolio positions. The
sponsor chooses maintenance according to
\[
V^I(q;B)
\equiv
 s \gamma qB-\psi(q),
\]
where the superscript $I$ denotes the decentralized institution in which
the sponsor internalizes only its retained stake.

As a benchmark, consider the maintenance choice that would arise if all
current token holders jointly controlled the decision. The corresponding
objective is
\[
V^A(q;B)
\equiv
\gamma qB-\psi(q),
\]
where the superscript $A$ denotes the all-owner benchmark. This is not a
social welfare function. It is the joint continuation value of current token
holders at the post-trade stage, taking the price already paid and the future
resale value as given.

The difference between the two problems is the value of the outside float.
Differentiating the two objectives gives

\[
\frac{\partial V^A(q;B)}{\partial q}
-
\frac{\partial V^I(q;B)}{\partial q}
=
(1- s )\gamma B\geq0 .
\]
The inequality is strict when $ s <1$ and $B>0$.
The wedge is the marginal continuation value generated by the passive
holdings. The sponsor protects the resale value of the entire token, but it
captures only the fraction $ s $ that it retains. Because the maintenance
cost is independent of ownership concentration,
the sponsor bears the full cost while internalizing only part of the benefit.

Define the all-owner maintenance choice as

\[
q^A(B)
\equiv
\arg\max_{q\in[0,1)}
\{\gamma qB-\psi(q)\}.
\]
The corresponding value function is
\[
V^A(B)
=
\psi^*(\gamma B).
\]

The decentralized sponsor chooses
\[
\mathbf{q}(B)
\equiv
\arg\max_{q\in[0,1)}
\{ s \gamma qB-\psi(q)\},
\]
with value
\[
V^I(B)
=
\psi^*( s \gamma B).
\]

\begin{proposition}[Governance wedge]
\label{prop:governance_wedge}

Fix a continuation price $B>0$. If $ s <1$, then

\[
\mathbf{q}(B)\leq q^A(B)
\]

and

\[
V^A(B)-V^I(B)
=
\psi^*(\gamma B)
-
\psi^*( s \gamma B)
\geq0 .
\]
Both inequalities are strict whenever $\gamma B>\psi'(0)$. In particular,
if $\psi'(0)=0$, they are strict for every $B>0$.

\end{proposition}

The proposition isolates the governance distortion created by incomplete
ownership retention. The sponsor controls maintenance but does not
internalize the value accruing to the outside float. In primal terms, once
the all-owner maintenance margin is active, the distortion appears as
insufficient maintenance,
\[
\mathbf{q}(B)<q^A(B);
\]
in dual terms, it appears as the value gap
\[
\psi^*(\gamma B)-\psi^*( s \gamma B).
\]

A larger retained stake narrows the incentive wedge and moves the
maintenance choice toward the all-owner benchmark, without creating a
fundamental payoff for the token. The two optimized value functions use
different ownership weights. Their difference is therefore not, by itself,
a measure of deadweight loss or a welfare comparison across generations.

\section{Nonlinear old-age utility}
\label{sec:general_uv}

This appendix allows nonlinear old-age utility while retaining the
baseline endowments and timing.
Assumption~\ref{ass:general_uv_preferences} replaces
Assumption~\ref{ass:linear_v_utility}; the maintenance technology in
Assumption~\ref{ass:maintenance} is unchanged.
The symbols $\underline q$, $\overline B$, $\mathcal{B}$, and
$ s _{\min}$ are redefined below for this extension. Its proofs are
collected in Appendix~\ref{app:proofs_general}.

Each generation is endowed with $e_1>0$ units of the consumption
good when young and $e_2\geq0$ units when old. Preferences are
\[
    u(c^y_{it})+\gamma E_t\!\left[v(c^o_{i,t+1})\right]
    -\text{maintenance disutility, if any},
\]
where $\gamma>0$. As in the baseline, we consider deterministic
price sequences conditional on survival. The survival realization
is the only aggregate uncertainty.

\begin{assumption}[Preferences]
\label{ass:general_uv_preferences}
The function $u:(0,e_1]\to\mathbb{R}$ is twice continuously
differentiable and satisfies
\[
    u'>0,\qquad u''<0,\qquad
    \lim_{c\downarrow0}u'(c)=+\infty.
\]
The function $v:[e_2,\infty)\to\mathbb{R}$ is continuously
differentiable, with a right derivative at $e_2$, and is twice
continuously differentiable on $(e_2,\infty)$. It satisfies
\[
    v'>0, \qquad
    v''\leq0.
\]
In particular, $v(e_2)$ and the right derivative $v'(e_2)$ are finite.
Moreover,
\begin{equation}
    u'(e_1)<\gamma v'(e_2).
    \label{eq:general_dynamic_inefficiency}
\end{equation}
\end{assumption}

Condition \eqref{eq:general_dynamic_inefficiency} is the generalized
dynamic-inefficiency restriction. Strict concavity of $u$ ensures
unique individual demand and a strictly increasing stationary
bubble price.

\subsection{Household demand}

At a live date $t$, a young agent purchasing $a_{it}$ tokens consumes
\[
    c^y_{it}=e_1-B_ta_{it},
    \qquad
    c^o_{i,t+1}=
    \begin{cases}
        e_2+B_{t+1}a_{it}, & Z_{t+1}=1,\\
        e_2, & Z_{t+1}=0.
    \end{cases}
\]
Taking prices and the anticipated survival probability as given,
the agent solves
\[
    \max_{0\leq a_{it}<e_1/B_t}
    \left\{
        u(e_1-B_ta_{it})
        +\gamma\left[
            q_t v(e_2+B_{t+1}a_{it})+(1-q_t)v(e_2)
        \right]
    \right\}.
\]
As in the baseline, maintenance disutility is additive and does
not change with an individual atomistic agent's portfolio choice,
so it does not enter this problem.

The objective is strictly concave. At an interior optimum,
\[
    B_tu'(e_1-B_ta_{it})
    =\gamma q_tB_{t+1}v'(e_2+B_{t+1}a_{it}).
\]
Symmetry and market clearing give $a_{it}=1$, and therefore
\[
    B_tu'(e_1-B_t)
    =\gamma q_tB_{t+1}v'(e_2+B_{t+1}).
\]
A positive live price must satisfy $B_t<e_1$. Aggregate consumption
at a live date is $(e_1-B_t)+(e_2+B_t)=e_1+e_2$, so token payments
remain transfers between generations.

\subsection{The stationary bubble price}

In a stationary live state, $B_t=B_{t+1}=B>0$, and household
optimality becomes
\begin{equation}
    u'(e_1-B)=\gamma qv'(e_2+B).
    \label{eq:general_stationary_pricing_condition}
\end{equation}
Define
\[
    \underline q\equiv
    \frac{u'(e_1)}{\gamma v'(e_2)}\in(0,1),
\]
and let $\overline B\in(0,e_1)$ solve
\[
    u'(e_1-\overline B)=\gamma v'(e_2+\overline B).
\]
For $q\in(\underline q,1)$, denote the solution to
\eqref{eq:general_stationary_pricing_condition} by $\mathcal{B}(q)$.

\begin{lemma}[Generalized Weil restriction]
\label{lem:general_weil}
For a given $q\in[0,1)$, a positive stationary household price
exists if and only if $q>\underline q$. It is uniquely given by
$B=\mathcal{B}(q)\in(0,\overline B)$. The bubble price is
continuously differentiable and strictly increasing, with
\begin{equation}
    \mathcal{B}'(q)
    =\frac{
        \gamma v'(e_2+\mathcal{B}(q))
    }{
        -u''(e_1-\mathcal{B}(q))
        -\gamma qv''(e_2+\mathcal{B}(q))
    }>0,
    \label{eq:general_pricing_locus_derivative}
\end{equation}
and endpoint limits
\[
    \lim_{q\downarrow\underline q}\mathcal{B}(q)=0,
    \qquad
    \lim_{q\uparrow1}\mathcal{B}(q)=\overline B.
\]
\end{lemma}

The threshold $\underline q$ is the infimum of the survival probabilities
consistent with a positive stationary price. The inequality must be strict. The bound
$\overline B$ is the limiting stationary price as survival tends
to certainty. These are stationary restrictions, not bounds on
all nonstationary prices or survival probabilities.

\subsection{The sponsor's post-trade problem}

For a continuation price $B\in[0,e_1]$, define the survival utility
gain
\[
    W(B)\equiv v(e_2+B)-v(e_2).
\]
The manager maximizes the aggregate utility of the
mass $ s $ of sponsor members. Omitting terms independent of
maintenance, the date-$t$ problem is
\[
    \max_{q\in[0,1)}
    \left\{ s \gamma qW(B_{t+1})-\psi(q)\right\}.
\]
Thus the best response to a continuation price $B$ is
\[
    \mathbf{q}(B)=
    \begin{cases}
        0,
        &  s \gamma W(B)\leq\psi'(0),\\[1mm]
        (\psi')^{-1}\!\left( s \gamma W(B)\right),
        &  s \gamma W(B)>\psi'(0).
    \end{cases}
\]
This function is continuous and weakly increasing. Wherever
$\mathbf{q}(B)>0$, it is continuously differentiable, with
\begin{equation}
    \mathbf{q}'(B)
    =\frac{ s \gamma v'(e_2+B)}{\psi''(\mathbf{q}(B))}>0.
    \label{eq:general_sponsor_derivative}
\end{equation}
Along a positive live path, household pricing requires $q_t>0$.
The sponsor's choice is therefore interior and satisfies
\[
    q_t=\mathbf{q}(B_{t+1}),
    \qquad
    \psi'(q_t)= s \gamma W(B_{t+1}).
\]

\subsection{Stationary equilibrium}

\begin{theorem}[Existence and minimum retention]
\label{thm:general_existence}
Under Assumptions~\ref{ass:general_uv_preferences}
and~\ref{ass:maintenance}, a positive stationary live
equilibrium exists if and only if $ s \geq s _{\min}$, where
\[
     s _{\min}
    \equiv
    \min_{q\in(\underline q,1)}
    \frac{\psi'(q)}{\gamma W(\mathcal{B}(q))}>0.
\]
A feasible retention level $ s \in(0,1]$ supports a positive
stationary price if and only if $ s _{\min}\leq1$.
\end{theorem}

The threshold compares marginal maintenance cost with the utility
value of survival supported by household demand. The bubbleless
stationary equilibrium also persists: if the continuation price is
zero, the sponsor chooses zero maintenance, and no household is
willing to pay a positive current price.

The following restriction replaces the baseline curvature condition in
Assumption~\ref{ass:linear_v_curvature} and adds a sufficient condition
for a single-valued, increasing dynamic price transition.

\Needspace{11\baselineskip}
\begin{assumption}[Curvature and transition monotonicity]
\label{ass:general_demand_geometry}
The marginal maintenance cost $\psi'$ is weakly convex. In addition,
\begin{equation}
    q\longmapsto W(\mathcal{B}(q))
    \quad\text{is weakly concave on }(\underline q,1),
    \label{eq:general_stationary_value_concavity}
\end{equation}
and
\begin{equation}
    v'(e_2+B)+Bv''(e_2+B)\geq0,
    \qquad B\in(0,e_1].
    \label{eq:general_C_increasing}
\end{equation}
\end{assumption}

Condition~\eqref{eq:general_C_increasing} is a sufficient condition
for a monotone price transition. It is not needed for stationary
existence or multiplicity.\footnote{Condition~\eqref{eq:general_stationary_value_concavity} can also be
checked directly from preferences: it is equivalent to requiring
\[
    B\longmapsto
    \frac{
        -u''(e_1-B)v'(e_2+B)
        -u'(e_1-B)v''(e_2+B)
    }{
        [v'(e_2+B)]^3
    }
\]
to be weakly increasing on $(0,\overline B)$. This equivalence
follows by differentiating $W(\mathcal{B}(q))$ and using the
stationary pricing condition. It requires only the maintained
second derivatives, not third derivatives.}

When $v(c)=c$, condition~\eqref{eq:general_C_increasing} is
automatic and the stationary bubble price becomes
\[
    \mathcal{B}(q)=e_1-(u')^{-1}(\gamma q).
\]
In this case, convexity of $u'$ implies
\eqref{eq:general_stationary_value_concavity}, recovering the
linear-old-age-utility assumptions.

\begin{proposition}[Multiplicity]
\label{prop:general_multiplicity}
Under Assumptions~\ref{ass:general_uv_preferences},
\ref{ass:maintenance}, and~\ref{ass:general_demand_geometry}, if
$ s > s _{\min}$, there are exactly two positive stationary
equilibria $(B_-,q_-)$ and $(B_+,q_+)$, with
\[
    \underline q<q_-<q_+<1,
    \qquad
    0<B_-<B_+<\overline B.
\]
\end{proposition}

\subsection{Dynamics of the bubble price}
\label{sec:general_forward_dynamics}

Substituting the sponsor best response into household pricing gives
\begin{equation}
    B_tu'(e_1-B_t)
    =\gamma\mathbf{q}(B_{t+1})B_{t+1}v'(e_2+B_{t+1}).
    \label{eq:general_forward_equation}
\end{equation}
For each current price admitting a solution $B_{t+1}\in(0,e_1)$,
this equation defines the live-price transition
\[
    B_{t+1}=T(B_t).
\]
The map is defined only where a positive feasible continuation
exists. No extension beyond that domain is required.

\begin{theorem}[Dynamics of the bubble price]
\label{thm:general_forward_dynamics}
Suppose Assumptions~\ref{ass:general_uv_preferences},
\ref{ass:maintenance},
and~\ref{ass:general_demand_geometry} hold, and
$ s > s _{\min}$.
Conditional on survival, every $B_0\in(0,B_+)$ generates a feasible
price path converging monotonically to $B_-$. The price $B_0=B_+$
generates the upper stationary path. For $B_0\in(B_+,e_1)$, a positive
feasible continuation either fails to exist immediately or fails after
finitely many iterations. The feasible positive initial prices are
therefore exactly $(0,B_+]$.
\end{theorem}

 The zero-price equilibrium is treated separately from
the positive-price transition. In particular, the proof does not
assume $T(0)=0$ or $\psi'(0)=0$.

\clearpage
\section{Proofs}
\label{app:proofs}

\subsection{The baseline model}
\label{app:proofs_baseline}

The following proofs use the baseline model with linear old-age utility.

We record some properties of the baseline best response
$\mathbf{q}$, including its zero-maintenance corner. In the following
lemma, $B$ denotes the continuation price, and the dependence of
$\mathbf{q}$ on $ s $ is suppressed in the notation.

\begin{lemma}[Properties of the sponsor best response]
\label{lem:qD}

The sponsor best response defined
in \eqref{eq:qD_general} satisfies

\[
\mathbf{q}(0)=0,\quad
\lim_{ s \rightarrow0}\mathbf{q}(B)=0,
\quad
\mathbf{q}(B)<1,\;\text{for any finite $B$.}
\]
Moreover, whenever the solution is interior,

\[
\frac{\partial \mathbf{q}(B)}{\partial B}>0,
\quad
\frac{\partial \mathbf{q}(B)}{\partial s }>0.
\]

\end{lemma}

\begin{proof}[Proof of Lemma~\ref{lem:linear_v_weil}]
Because $u'$ is continuous and strictly decreasing, the stationary
condition \eqref{eq:linear_v_stationary_pricing_condition} has a
positive-price solution if and only if $\gamma q>u'(e_1)$, that is
$q>\underline q$. The Inada condition then gives the unique solution
\eqref{eq:linear_v_pricing_locus}. Since $(u')^{-1}$ is strictly
decreasing, $\mathcal{B}$ is strictly increasing. Thus, the endpoint limits follow
from \eqref{eq:linear_v_bounds}.
\end{proof}

\begin{proof}[Proof of Theorem~\ref{thm:linear_v_existence}]
A positive stationary equilibrium requires $B=\mathcal{B}(q)$ and
$\psi'(q)= s \gamma \mathcal{B}(q)$, or equivalently $ s =h(q)$ where
\[
    h(q)
    \equiv
    \frac{\psi'(q)}{\gamma [e_1-(u')^{-1}(\gamma q)]},
    \qquad
    q\in(\underline q,1).
\]
The function $h$ is continuous. As $q\downarrow\underline q$ the
denominator converges to zero from above while
$\psi'(q)\to\psi'(\underline q)>0$, so $h(q)\to+\infty$. As
$q\uparrow1$ the denominator converges to the finite positive number
$\gamma\overline B$ while $\psi'(q)\to+\infty$, so again
$h(q)\to+\infty$. Hence $h$ attains a strictly positive minimum at an
interior point, and a solution to $h(q)= s $ exists if and only if
$ s $ is weakly above this minimum. The feasibility statement
follows because $ s $ is a share and therefore cannot exceed one.
\end{proof}

\begin{proof}[Proof of Proposition~\ref{prop:linear_v_multiplicity}]
Since $u'$ is decreasing and convex, its inverse is decreasing and
convex. Hence $\mathcal{B}(q)=e_1-(u')^{-1}(\gamma q)$ is concave
under Assumption~\ref{ass:linear_v_curvature}. Define the \emph{equilibrium gap}
\begin{equation}
    G(q; s )
    \equiv
     s \gamma\mathcal{B}(q)-\psi'(q),
    \qquad
    q\in(\underline q,1).
    \label{eq:linear_v_gap}
\end{equation}
A positive stationary equilibrium is exactly a zero of $G(\cdot; s )$. Since $\psi'$ is
convex, $-\psi'$ is concave, and therefore $G(\cdot; s )$ is
concave. Moreover
\[
    \lim_{q\downarrow\underline q}G(q; s )
    =
    -\psi'(\underline q)<0,
    \qquad
    \lim_{q\uparrow1}G(q; s )
    =
    -\infty.
\]
When $ s > s _{\min}$, $G$ is strictly positive at some interior
point. Continuity gives one zero on each side of the positive region.
Concavity implies that the nonnegative set of $G$ is an interval and
rules out any additional crossing. Hence there are exactly two zeros
$q_-<q_+$, with
\[
    G(q; s )
    \begin{cases}
        <0, & q\in(\underline q,q_-),\\
        >0, & q\in(q_-,q_+),\\
        <0, & q\in(q_+,1).
    \end{cases}
\]
Concavity and strict positivity between the roots also imply
\[
    G_q(q_-; s )>0>G_q(q_+; s ).
\]
Since $\mathcal{B}$ is strictly increasing, $B_-<B_+$.
\end{proof}

\begin{proof}[Proof of Lemma~\ref{lem:comparative}]
Making the dependence on $\gamma$ explicit, the equilibrium gap in
\eqref{eq:linear_v_gap} is
\[
    G(q; s ,\gamma)= s \gamma\mathcal{B}(q;\gamma)-\psi'(q),
    \qquad
    \mathcal{B}(q;\gamma)=e_1-(u')^{-1}(\gamma q).
\]
The proof of Proposition~\ref{prop:linear_v_multiplicity} gives
$G_q(q_-; s ,\gamma)>0>G_q(q_+; s ,\gamma)$.
At either equilibrium, implicit differentiation therefore yields
\[
    \frac{\partial q_i}{\partial s }
    =-\frac{G_ s (q_i; s ,\gamma)}{G_q(q_i; s ,\gamma)},
    \qquad
    \frac{\partial q_i}{\partial\gamma}
    =-\frac{G_\gamma(q_i; s ,\gamma)}{G_q(q_i; s ,\gamma)},
    \qquad i\in\{-,+\}.
\]
Both partial derivatives in the numerators are positive, since
\[
    G_ s =\gamma\mathcal{B}>0,
    \qquad
    G_\gamma= s \left[
        \mathcal{B}+\gamma\frac{\partial\mathcal{B}}{\partial\gamma}
    \right]>0,
    \qquad
    \frac{\partial\mathcal{B}}{\partial\gamma}
    =-\frac{q}{u''(e_1-\mathcal{B})}>0.
\]
This establishes the signs of both comparative statics on each survival branch.

Because $\mathcal{B}$ has no direct dependence on $ s $ and
$\mathcal{B}_q=-\gamma/u''>0$, the signs of
$\partial B_i/\partial s $ coincide with those of
$\partial q_i/\partial s $.
For $\gamma$, the direct effect on household pricing must also be
included. Using the preceding derivatives and
$G_q= s \gamma\mathcal{B}_q-\psi''$, we obtain
\begin{align*}
    \frac{\partial B_i}{\partial\gamma}
    &=\frac{\partial\mathcal{B}}{\partial\gamma}(q_i;\gamma)
      +\frac{\partial\mathcal{B}}{\partial q}(q_i;\gamma)
       \frac{\partial q_i}{\partial\gamma}\\
    &=\frac{ s \gamma B_i+q_i\psi''(q_i)}
    {u''(e_1-B_i)G_q(q_i; s ,\gamma)}.
\end{align*}
The numerator is positive and $u''<0$, so this derivative has the
opposite sign to $G_q(q_i; s ,\gamma)$. Thus
$\partial B_+/\partial\gamma>0$ and
$\partial B_-/\partial\gamma<0$, as claimed.
\end{proof}

\begin{proof}[Proof of Theorem~\ref{thm:linear_v_forward_dynamics}]
For each current price $B\in(0,e_1)$, the associated maintenance
choice $q\in(0,1)$ and the continuation price satisfy
\[
    q\psi'(q)= s  B u'(e_1-B),
    \qquad
    T(B)=\frac{\psi'(q)}{ s \gamma}.
\]
Both $B\mapsto Bu'(e_1-B)$ and $q\mapsto q\psi'(q)$ are
continuous and strictly increasing, with range $(0,\infty)$.
Thus $q$ is uniquely determined and strictly increasing in $B$.
Implicit differentiation gives
\[
    T'(B)
    =
    \frac{
        \psi''(q)\bigl[u'(e_1-B)-Bu''(e_1-B)\bigr]
    }{
        \gamma\bigl[\psi'(q)+q\psi''(q)\bigr]
    }
    >0.
\]
Hence $T$ is continuously differentiable and strictly
increasing on $(0,e_1)$.

A positive fixed point $T(B)=B$ satisfies
\[
    \psi'(q)= s \gamma B,
    \qquad
    u'(e_1-B)=\gamma q,
\]
which are exactly the stationary equilibrium conditions.
Conversely, every positive stationary equilibrium is a fixed
point. Proposition~\ref{prop:linear_v_multiplicity} therefore implies that the only positive
fixed points are $B_-$ and $B_+$.

We next determine the direction of motion.
At either stationary equilibrium $(B_i,q_i)$, $i\in\{-,+\}$,
differentiating the equilibrium gap in \eqref{eq:linear_v_gap} gives
\[
    G_q(q_i; s )
    =
    -\frac{ s \gamma^2}{u''(e_1-B_i)}
    -\psi''(q_i).
\]
Using the stationary conditions in the expression for
$T'$ yields
\[
    T'(B_i)-1
    =
    \frac{
        B_i u''(e_1-B_i)
    }{
        \gamma\bigl[\psi'(q_i)+q_i\psi''(q_i)\bigr]
    }
    G_q(q_i; s ).
\]
The coefficient multiplying $G_q$ is strictly negative.
By Proposition~\ref{prop:linear_v_multiplicity}, $G$ is concave and strictly positive between
its two zeros, so
\[
    G_q(q_-; s )>0>G_q(q_+; s ).
\]
Consequently,
\[
    0<T'(B_-)<1<T'(B_+).
\]

These slope inequalities determine the sign of
$T(B)-B$ near each fixed point.
Since this difference is continuous and has no other zeros,
its sign cannot change elsewhere within any of the three
intervals. Therefore,
\[
    T(B)-B
    \begin{cases}
        >0, & B\in(0,B_-),\\
        <0, & B\in(B_-,B_+),\\
        >0, & B\in(B_+,e_1).
    \end{cases}
\]

Suppose first that $B_0\in(0,B_-)$.
The sign pattern and strict monotonicity of $T$ imply
\[
    B_t<B_{t+1}<B_-
\]
at every date along the live path.
The sequence is therefore increasing and bounded above by
$B_-$. Its limit is positive and, by continuity, must be a
fixed point. Hence $B_t\uparrow B_-$.

Similarly, if $B_0\in(B_-,B_+)$, then
\[
    B_-<B_{t+1}<B_t,
\]
so the sequence decreases to a fixed point in $[B_-,B_0]$.
The only such fixed point is $B_-$, and therefore
$B_t\downarrow B_-$.
The cases $B_0=B_-$ and $B_0=B_+$ yield stationary paths.

Finally, suppose $B_0\in(B_+,e_1)$.
The implied price sequence is strictly increasing for as long
as it remains feasible. Suppose, toward a contradiction, that
$B_t<e_1$ at every date. Then $B_t$ converges to some
$\ell\in(B_+,e_1]$.
If $\ell<e_1$, continuity implies
\[
    T(\ell)=\ell,
\]
contradicting the absence of a fixed point above $B_+$.
If $\ell=e_1$, the household pricing condition implies
\[
    B_tu'(e_1-B_t)
    =
    \gamma q_tB_{t+1}
    <
    \gamma e_1
\]
at every date. This contradicts
$u'(e_1-B_t)\to+\infty$ as $B_t\uparrow e_1$.

Thus the implied continuation price must reach or exceed
$e_1$ after finitely many iterations, violating feasibility.
No initial price above $B_+$ can therefore be part of an
infinite-horizon positive live equilibrium.
\end{proof}

The same geometry delivers branch-specific comparative statics.

\begin{lemma}[Comparative statics]
\label{lem:comparative}
Suppose Assumptions~\ref{ass:linear_v_utility}, \ref{ass:maintenance}, and~\ref{ass:linear_v_curvature} hold and $ s > s _{\min}$.
\[
\frac{\partial B_+}{\partial s }>0,
\qquad
\frac{\partial B_-}{\partial s }<0,
\qquad
\frac{\partial B_+}{\partial\gamma}>0,
\qquad
\frac{\partial B_-}{\partial\gamma}<0 .
\]
The same signs apply to the corresponding survival probabilities $q_\pm$.
\end{lemma}

Greater retention moves the stationary equilibria in opposite directions.
On the upper branch, price and survival increase. On the lower branch,
both decrease. A larger old-age utility weight $\gamma$ has the same
sign pattern. 

\begin{proof}[Proof of Corollary~\ref{cor:viable_range}]
Fix $ s > s _{\min}$. By
Theorem~\ref{thm:linear_v_forward_dynamics}, every initial price in
$(0,B_+( s ))$ has a feasible path converging to
$B_-( s )$, and $B_+( s )$ has the upper stationary continuation.
Every larger initial price eventually violates feasibility. Hence
$\mathcal{V}( s )=(0,B_+( s )]$.
Lemma~\ref{lem:comparative} gives
$\partial B_+/\partial s >0$. Therefore
$ s _{\min}< s '< s \leq1$ implies
\[
(0,B_+( s ')]\subsetneq(0,B_+( s )],
\]
as claimed.
\end{proof}

\begin{proof}[Proof of Theorem~\ref{thm:retention_crash}]
Write $q_r$ for the maintenance choice associated with a
current price $B\in(0,e_1)$ and a positive retained share $r$, defined by
\[
    q_r\psi'(q_r)
    =r B u'(e_1-B).
\]
The dependence of $q_r$ on the fixed current price $B$ is suppressed.
Because $q\mapsto q\psi'(q)$ is continuous and strictly increasing
from zero to infinity, $q_r\in(0,1)$ is uniquely defined and
strictly increasing in $r$. The candidate continuation is
\[
    T_r(B)=\frac{\psi'(q_r)}{r\gamma}
          =\frac{B u'(e_1-B)}{\gamma q_r}.
\]
Thus $T_r(B)$ is strictly decreasing in $r$. In particular, for $0<r\leq\widehat s$,
\[
    T_r(B)\geq T_{\widehat s}(B).
\]

The map $T_{\widehat s}$ is continuous on $(0,e_1)$ and tends to
infinity as $B\uparrow e_1$. A positive fixed point would be a stationary
equilibrium with retention $\widehat s$, which is impossible by
Theorem~\ref{thm:linear_v_existence} because $\widehat s<s_{\min}$.
Continuity and the behavior near $e_1$ therefore imply
\[
    T_{\widehat s}(B)>B
    \qquad\text{for every }B\in(0,e_1).
\]

Suppose a positive equilibrium price could be sustained at date $t_*$.
Household pricing requires a positive survival probability and a positive
resale value at every subsequent live date. If $s_n=0$ at any such date,
the sponsor chooses zero maintenance, immediately contradicting this
requirement. Otherwise $s_n\in(0,\widehat s]$ for every $n\geq t_*$,
and equilibrium requires
\[
    B_{n+1}=T_{s_n}(B_n)
    \geq T_{\widehat s}(B_n)>B_n.
\]
The positive price sequence is therefore strictly increasing. Household
pricing and feasibility also imply
\[
    B_nu'(e_1-B_n)=\gamma q_nB_{n+1}<\gamma e_1.
\]
The Inada condition bounds $B_n$ away from $e_1$, so the sequence
converges to a limit $\ell\in(0,e_1)$. Passing to the limit in the
preceding transition inequality gives
\[
    \ell\geq T_{\widehat s}(\ell)>\ell,
\]
a contradiction. There is no positive infinite-horizon continuation
from date $t_*$, and hence $B_{t_*}=0$ on any history reaching that date.

At date $\tau$, this absence of a positive continuation is common
knowledge. If $t_*>\tau$, a positive price at $t_*-1$ would require a
positive resale value at $t_*$, which has just been ruled out. Repeating
the argument backward yields
\[
    B_n=0,\qquad n=\tau,\ldots,t_*.
\]
In particular, $B_\tau=0$. This conclusion concerns prices after the
announcement; it does not alter payments made before the information
became available.
\end{proof}

\subsection{Additional properties of the sponsor best response}
\label{app:proofs_sponsor}

\begin{proof}[Proof of Lemma~\ref{lem:qD}]
For a fixed continuation price $B\geq0$, the sponsor's objective has
first derivative $ s \gamma B-\psi'(q)$ and second derivative
$-\psi''(q)<0$. If $ s \gamma B\leq\psi'(0)$, its derivative is
nonpositive at zero and strictly negative thereafter, so zero is the
unique maximizer. Otherwise, the derivative is positive at zero and tends
to $-\infty$ as $q\uparrow1$, giving the unique interior solution in
\eqref{eq:qD_general}. In particular, $\mathbf{q}(0)=0$, and every
finite $B$ gives $\mathbf{q}(B)<1$.

As $ s \downarrow0$, the marginal benefit $ s \gamma B$ tends
to zero. If $\psi'(0)>0$, the choice is eventually the zero corner.
If $\psi'(0)=0$, continuity and strict monotonicity of $\psi'$ imply
$(\psi')^{-1}( s \gamma B)\to0$. Hence
$\mathbf{q}(B)\to0$ in either case.

At an interior solution, implicit differentiation of
$\psi'(\mathbf{q}(B))= s \gamma B$ gives
\[
    \frac{\partial\mathbf{q}(B)}{\partial B}
    =\frac{ s \gamma}{\psi''(\mathbf{q}(B))}>0,
    \qquad
    \frac{\partial\mathbf{q}(B)}{\partial s }
    =\frac{\gamma B}{\psi''(\mathbf{q}(B))}>0.
\]
This proves the stated properties.
\end{proof}

\subsection{Nonlinear old-age utility}
\label{app:proofs_general}

All objects in this subsection are those of Appendix~\ref{sec:general_uv}.

\begin{proof}[Proof of Lemma~\ref{lem:general_weil}]
For a fixed $q\geq0$, the difference
\[
    u'(e_1-B)-\gamma qv'(e_2+B)
\]
is strictly increasing in $B$ and diverges to $+\infty$ as
$B\uparrow e_1$. Its value at $B=0$ is negative exactly when
$q>\underline q$. This proves existence and uniqueness, including
the existence of $\overline B$ by setting $q=1$.
Implicit differentiation gives
\eqref{eq:general_pricing_locus_derivative}. Monotonicity and
continuity give the endpoint limits.
\end{proof}

\begin{proof}[Proof of Theorem~\ref{thm:general_existence}]
A positive stationary equilibrium satisfies
$B=\mathcal{B}(q)$ and $q=\mathbf{q}(B)$. Equivalently, it is a
zero of the equilibrium gap
\begin{equation}
    G(q; s )
    \equiv s \gamma W(\mathcal{B}(q))-\psi'(q),
    \qquad q\in(\underline q,1).
    \label{eq:general_equilibrium_gap}
\end{equation}
The stationary conditions are equivalent to
\[
     s =
    \frac{\psi'(q)}{\gamma W(\mathcal{B}(q))}.
\]
The ratio is positive and continuous on $(\underline q,1)$.
As $q\downarrow\underline q$, its denominator tends to zero while
its numerator tends to $\psi'(\underline q)>0$.
As $q\uparrow1$, its denominator tends to the finite positive
value $\gamma W(\overline B)$ while its numerator diverges.
Hence the ratio diverges at both endpoints, attains a positive
minimum in the interior, and has range
$[ s _{\min},\infty)$. This proves both claims.
\end{proof}

\begin{proof}[Proof of Proposition~\ref{prop:general_multiplicity}]
The gap $G$ defined in \eqref{eq:general_equilibrium_gap} is concave, with
\[
    \lim_{q\downarrow\underline q}G(q; s )
    =-\psi'(\underline q)<0,
    \qquad
    \lim_{q\uparrow1}G(q; s )=-\infty.
\]
Since $ s > s _{\min}$, it is strictly positive at an
interior point. Continuity gives a zero on either side of that
point, and concavity rules out additional zeros. Thus
\[
    G(q; s )
    \begin{cases}
        <0, & q\in(\underline q,q_-),\\
        >0, & q\in(q_-,q_+),\\
        <0, & q\in(q_+,1).
    \end{cases}
\]
Moreover, concavity and strict positivity between the roots imply
\begin{equation}
    G_q(q_-; s )>0>G_q(q_+; s ).
    \label{eq:general_gap_slopes}
\end{equation}
The ordering of prices follows from $\mathcal{B}'>0$.
\end{proof}

\begin{proof}[Proof of Theorem~\ref{thm:general_forward_dynamics}]
First consider the right-hand side of
\eqref{eq:general_forward_equation} as a function of the
continuation price $x$. It is continuous, equals zero at $x=0$,
and is strictly increasing wherever $\mathbf{q}(x)>0$, because
\begin{align*}
    \frac{d}{dx}
    \left[\gamma\mathbf{q}(x)xv'(e_2+x)\right]
    ={}&\gamma\mathbf{q}'(x)xv'(e_2+x)\\
    &+\gamma\mathbf{q}(x)
    \left[v'(e_2+x)+xv''(e_2+x)\right]>0.
\end{align*}
Here the first term is strictly positive and the second is
nonnegative. By contrast, the current-price expression satisfies
\[
    \frac{d}{dB}\left[Bu'(e_1-B)\right]
    =u'(e_1-B)-Bu''(e_1-B)>0,
\]
and increases from zero to infinity on $[0,e_1)$.
Consequently, current prices admitting a positive feasible
continuation form an interval, and $T$ is uniquely defined,
continuously differentiable, and strictly increasing there.
This interval contains $(0,B_+]$ and some prices above $B_+$:
$B_+$ is a feasible stationary price, and the continuation-value
expression is strictly increasing in a neighborhood of $x=B_+$.

A positive fixed point satisfies
\[
    u'(e_1-B)=\gamma qv'(e_2+B),
    \qquad
    q=\mathbf{q}(B)>0.
\]
These are precisely the stationary equilibrium conditions.
Thus Proposition~\ref{prop:general_multiplicity} implies that
$B_-$ and $B_+$ are the only positive fixed points.

We next establish local stability and use it to determine the
global direction of motion. At a fixed point $(B_i,q_i)$,
$i\in\{-,+\}$, implicit differentiation gives
\begin{align*}
    u'(e_1-B_i)-B_i u''(e_1-B_i)
    ={}&\gamma T'(B_i)\Bigl\{
        B_i v'(e_2+B_i)\mathbf{q}'(B_i)\\
    &\qquad
        +q_i\left[v'(e_2+B_i)+B_i v''(e_2+B_i)\right]
    \Bigr\}.
\end{align*}
The coefficient of $T'(B_i)$ is strictly positive. Subtracting
that coefficient from the left-hand side, and using stationary
pricing together with
\eqref{eq:general_pricing_locus_derivative}, yields
\[
    \operatorname{sgn}\!\left[T'(B_i)-1\right]
    =\operatorname{sgn}\!\left[
        \frac{1}{\mathcal{B}'(q_i)}-\mathbf{q}'(B_i)
    \right].
\]
Meanwhile, differentiating \eqref{eq:general_equilibrium_gap} and using
\eqref{eq:general_sponsor_derivative} gives
\[
    G_q(q_i; s )
    =\psi''(q_i)
    \left[\mathbf{q}'(B_i)\mathcal{B}'(q_i)-1\right].
\]
Since $\mathcal{B}'>0$ and $\psi''>0$, it follows that
\[
    \operatorname{sgn}\!\left[T'(B_i)-1\right]
    =-\operatorname{sgn}G_q(q_i; s ).
\]
Together with \eqref{eq:general_gap_slopes} and $T'>0$, this proves
\[
    0<T'(B_-)<1<T'(B_+).
\]

These slopes determine the sign of $T(B)-B$ near the two fixed
points. Since the difference is continuous and has no other
zeros on its interval of definition, its sign cannot change
elsewhere within the corresponding intervals. Hence
\begin{equation}
    T(B)-B
    \begin{cases}
        >0, & 0<B<B_-,\\
        <0, & B_-<B<B_+,\\
        >0, & B>B_+,
    \end{cases}
    \label{eq:general_forward_signs}
\end{equation}
where the last line applies only when a feasible continuation exists.

If $B_0\in(0,B_-)$, strict monotonicity and
\eqref{eq:general_forward_signs} give
$B_t<B_{t+1}<B_-$ at every date along the live path.
The sequence increases to a positive fixed point, necessarily
$B_-$. If $B_0\in(B_-,B_+)$, the same argument gives
$B_-<B_{t+1}<B_t$, so the sequence decreases to $B_-$.
The cases $B_0=B_-$ and $B_0=B_+$ are stationary.

Finally, let $B_0\in(B_+,e_1)$. If a feasible continuation exists,
it is strictly greater than the current price. Suppose, toward
a contradiction, that feasible continuations exist at every
date. Then the price sequence is increasing and bounded above
by $e_1$, and therefore converges to some $\ell\in(B_+,e_1]$.

If $\ell<e_1$, continuity of the sponsor best response implies
$q_t=\mathbf{q}(B_{t+1})\to\mathbf{q}(\ell)>0$.
Taking limits in household pricing gives
\[
    u'(e_1-\ell)
    =\gamma\mathbf{q}(\ell)v'(e_2+\ell).
\]
Thus $(\ell,\mathbf{q}(\ell))$ is a stationary equilibrium above
$B_+$, a contradiction. This argument uses the equilibrium
conditions directly and does not require $T$ to be defined at
a boundary of its domain.

If $\ell=e_1$, household pricing instead implies
\[
    B_tu'(e_1-B_t)
    =\gamma q_tB_{t+1}v'(e_2+B_{t+1})
    \leq\gamma e_1v'(e_2)<\infty,
\]
contradicting the Inada condition as $B_t\uparrow e_1$.
Therefore an initial price above $B_+$ either admits no feasible
continuation immediately or ceases to admit one after finitely
many iterations. Such a price cannot be part of an infinite-horizon
positive live equilibrium.
\end{proof}

\subsection{The governance externality}
\label{app:proofs_governance}

This proof returns to the baseline linear-old-age-utility comparison.

\begin{proof}[Proof of Proposition~\ref{prop:governance_wedge}]
The two optimization problems differ only in the coefficient on the
continuation value of survival.

Because
\[
 s \gamma B<\gamma B
\]
when $ s <1$, and because the objective has increasing differences in
$(q,\gamma B)$, the optimal maintenance choice is weakly increasing in
the shadow value of survival. Therefore,
\[
\mathbf{q}(B)\leq q^A(B).
\]
If $\gamma B>\psi'(0)$, the all-owner solution is positive. Strict
convexity of $\psi$ and the strictly smaller sponsor coefficient then imply
$\mathbf{q}(B)<q^A(B)$, whether the sponsor is at the zero corner or in the
interior.

For the value comparison, the Fenchel conjugate representation gives
\[
V^I(B)=\psi^*( s \gamma B),
\qquad
V^A(B)=\psi^*(\gamma B).
\]
Since the convex conjugate is weakly increasing in its argument,
\[
V^A(B)\geq V^I(B).
\]
It is strictly increasing once its optimizer is positive, which gives the
stated strict value comparison under $\gamma B>\psi'(0)$.
\end{proof}

\bibliographystyle{plainnat}
\bibliography{who_keeps_a_bubble_alive_refs}

@article{fieldhanka2001,
  author  = {Field, Laura Casares and Hanka, Gordon},
  title   = {The Expiration of {IPO} Share Lockups},
  journal = {Journal of Finance},
  volume  = {56},
  number  = {2},
  pages   = {471--500},
  year    = {2001}
}

@article{santoswoodford1997,
  author  = {Santos, Manuel S. and Woodford, Michael},
  title   = {Rational Asset Pricing Bubbles},
  journal = {Econometrica},
  year    = {1997},
  volume  = {65},
  number  = {1},
  pages   = {19--57},
  doi     = {10.2307/2171812}
}

@article{kocherlakota1992,
  author  = {Kocherlakota, Narayana R.},
  title   = {Bubbles and Constraints on Debt Accumulation},
  journal = {Journal of Economic Theory},
  year    = {1992},
  volume  = {57},
  number  = {1},
  pages   = {245--256},
  doi     = {10.1016/S0022-0531(05)80052-3}
}

@article{kocherlakota2008,
  author  = {Kocherlakota, Narayana},
  title   = {Injecting Rational Bubbles},
  journal = {Journal of Economic Theory},
  year    = {2008},
  volume  = {142},
  number  = {1},
  pages   = {218--232},
  doi     = {10.1016/j.jet.2006.07.010}
}

@article{hugonnier2012,
  author  = {Hugonnier, Julien},
  title   = {Rational Asset Pricing Bubbles and Portfolio Constraints},
  journal = {Journal of Economic Theory},
  year    = {2012},
  volume  = {147},
  number  = {6},
  pages   = {2260--2302},
  doi     = {10.1016/j.jet.2012.05.003}
}

@article{abreubrunnermeier2003,
  author  = {Abreu, Dilip and Brunnermeier, Markus K.},
  title   = {Bubbles and Crashes},
  journal = {Econometrica},
  year    = {2003},
  volume  = {71},
  number  = {1},
  pages   = {173--204},
  doi     = {10.1111/1468-0262.00393}
}

@article{chemla2014skin,
  author  = {Chemla, Gilles and Hennessy, Christopher A.},
  title   = {Skin in the Game and Moral Hazard},
  journal = {Journal of Finance},
  year    = {2014},
  volume  = {69},
  number  = {4},
  pages   = {1597--1641},
  doi     = {10.1111/jofi.12161}
}

@article{maug1998,
  author  = {Maug, Ernst},
  title   = {Large Shareholders as Monitors: Is There a Trade-Off between Liquidity and Control?},
  journal = {Journal of Finance},
  year    = {1998},
  volume  = {53},
  number  = {1},
  pages   = {65--98},
  doi     = {10.1111/0022-1082.35053}
}

@article{paganoroell1998,
  author  = {Pagano, Marco and R{\"o}ell, Ailsa},
  title   = {The Choice of Stock Ownership Structure: Agency Costs, Monitoring, and the Decision to Go Public},
  journal = {Quarterly Journal of Economics},
  year    = {1998},
  volume  = {113},
  number  = {1},
  pages   = {187--225},
  doi     = {10.1162/003355398555568}
}

@article{huvaras2025intermediary,
  author  = {Hu, Yunzhi and Varas, Felipe},
  title   = {Intermediary Financing without Commitment},
  journal = {Journal of Financial Economics},
  year    = {2025},
  volume  = {167},
  pages   = {104025},
  doi     = {10.1016/j.jfineco.2025.104025}
}

@article{pagnotta2022decentralizing,
  author  = {Pagnotta, Emiliano S.},
  title   = {Decentralizing Money: Bitcoin Prices and Blockchain Security},
  journal = {Review of Financial Studies},
  year    = {2022},
  volume  = {35},
  number  = {2},
  pages   = {866--907},
  doi     = {10.1093/rfs/hhaa149}
}

@article{conghe2019blockchain,
  author  = {Cong, Lin William and He, Zhiguo},
  title   = {Blockchain Disruption and Smart Contracts},
  journal = {Review of Financial Studies},
  year    = {2019},
  volume  = {32},
  number  = {5},
  pages   = {1754--1797},
  doi     = {10.1093/rfs/hhz007}
}

@article{congheli2021pools,
  author  = {Cong, Lin William and He, Zhiguo and Li, Jiasun},
  title   = {Decentralized Mining in Centralized Pools},
  journal = {Review of Financial Studies},
  year    = {2021},
  volume  = {34},
  number  = {3},
  pages   = {1191--1235},
  doi     = {10.1093/rfs/hhaa040}
}

@techreport{congstaking2025,
  author      = {Cong, Lin William and He, Zhiheng and Tang, Ke},
  title       = {The Tokenomics of Staking},
  institution = {National Bureau of Economic Research},
  type        = {Working Paper},
  number      = {33640},
  year        = {2025},
  doi         = {10.3386/w33640}
}

@article{canidio2023token,
  author  = {Canidio, Andrea},
  title   = {Token Distribution and Platform Adoption},
  journal = {AEA Papers and Proceedings},
  year    = {2023},
  volume  = {113},
  pages   = {263--267},
  doi     = {10.1257/pandp.20231032}
}

@article{sockinxiong2023tokenization,
  author  = {Sockin, Michael and Xiong, Wei},
  title   = {Decentralization through Tokenization},
  journal = {Journal of Finance},
  year    = {2023},
  volume  = {78},
  number  = {1},
  pages   = {247--299},
  doi     = {10.1111/jofi.13192}
}

@article{limann2025digital,
  author  = {Li, Jiasun and Mann, William},
  title   = {Digital Tokens and Platform Building},
  journal = {Review of Financial Studies},
  year    = {2025},
  volume  = {38},
  number  = {7},
  pages   = {1921--1954},
  doi     = {10.1093/rfs/hhaf015}
}

@article{ferreira2023corporate,
  author  = {Ferreira, Daniel and Li, Jin and Nikolowa, Radoslawa},
  title   = {Corporate Capture of Blockchain Governance},
  journal = {Review of Financial Studies},
  year    = {2023},
  volume  = {36},
  number  = {4},
  pages   = {1364--1407},
  doi     = {10.1093/rfs/hhac051}
}

@article{malinova2023tokenomics,
  author  = {Malinova, Katya and Park, Andreas},
  title   = {Tokenomics: When Tokens Beat Equity},
  journal = {Management Science},
  year    = {2023},
  volume  = {69},
  number  = {11},
  pages   = {6568--6583},
  doi     = {10.1287/mnsc.2023.4882}
}

@article{goldstein2025fragility,
  author  = {Goldstein, Itay and Huang, Chong and Yang, Liyan},
  title   = {Fragility of Financial Markets},
  journal = {Annual Review of Financial Economics},
  year    = {2025},
  volume  = {17},
  pages   = {27--48},
  doi     = {10.1146/annurev-financial-120522-114723}
}

@article{congliwangs2021tokenomics,
  author  = {Cong, Lin William and Li, Ye and Wang, Neng},
  title   = {Tokenomics: Dynamic Adoption and Valuation},
  journal = {The Review of Financial Studies},
  year    = {2021},
  volume  = {34},
  number  = {3},
  pages   = {1105--1155},
  doi     = {10.1093/rfs/hhaa089}
}

@article{congliwangs2022platform,
  author  = {Cong, Lin William and Li, Ye and Wang, Neng},
  title   = {Token-Based Platform Finance},
  journal = {Journal of Financial Economics},
  year    = {2022},
  volume  = {144},
  number  = {3},
  pages   = {972--991},
  doi     = {10.1016/j.jfineco.2021.10.002}
}

@article{congbiaisetal2023advances,
  author  = {Biais, Bruno and Capponi, Agostino and Cong, Lin William and Gaur, Vishal and Giesecke, Kay},
  title   = {Advances in Blockchain and Crypto Economics},
  journal = {Management Science},
  year    = {2023},
  volume  = {69},
  number  = {11},
  pages   = {6417--6426},
  doi     = {10.1287/mnsc.2023.intro.v69.n11}
}

@article{congwash2023,
  author  = {Cong, Lin William and Li, Xi and Tang, Ke and Yang, Yang},
  title   = {Crypto Wash Trading},
  journal = {Management Science},
  year    = {2023},
  volume  = {69},
  number  = {11},
  pages   = {6427--6454},
  doi     = {10.1287/mnsc.2021.02709}
}

@article{gantsoukalasnetessine2021ico,
  author  = {Gan, Jingxing and Tsoukalas, Gerry and Netessine, Serguei},
  title   = {Initial Coin Offerings, Speculation, and Asset Tokenization},
  journal = {Management Science},
  year    = {2021},
  volume  = {67},
  number  = {2},
  pages   = {914--931},
  doi     = {10.1287/mnsc.2020.3796}
}

@article{gantsoukalasnetessine2023dao,
  author  = {Gan, Jingxing and Tsoukalas, Gerry and Netessine, Serguei},
  title   = {Decentralized Platforms: Governance, Tokenomics, and {ICO} Design},
  journal = {Management Science},
  year    = {2023},
  volume  = {69},
  number  = {11},
  pages   = {6667--6683},
  doi     = {10.1287/mnsc.2021.02076}
}

@article{chodtrichakisyang2022,
  author  = {Chod, Jiri and Trichakis, Nikolaos and Yang, S. Alex},
  title   = {Platform Tokenization: Financing, Governance, and Moral Hazard},
  journal = {Management Science},
  year    = {2022},
  volume  = {68},
  number  = {9},
  pages   = {6411--6433},
  doi     = {10.1287/mnsc.2021.4225}
}

@article{hanleeli2026,
  author  = {Han, Jungsuk and Lee, Jongsub and Li, Tao},
  title   = {Decentralized Autonomous Organization Governance},
  journal = {Management Science},
  year    = {2026},
  note    = {Articles in Advance, published online June 12, 2026},
  doi     = {10.1287/mnsc.2024.07033}
}

@article{canidio2026financial,
  author  = {Canidio, Andrea},
  title   = {Financial Incentives for the Development of Blockchain-Based Platforms},
  journal = {Journal of Industrial Economics},
  year    = {2026},
  volume  = {74},
  number  = {1},
  pages   = {34--47},
  doi     = {10.1111/joie.70010}
}

@article{sockinxiong2023cryptocurrencies,
  author  = {Sockin, Michael and Xiong, Wei},
  title   = {A Model of Cryptocurrencies},
  journal = {Management Science},
  year    = {2023},
  volume  = {69},
  number  = {11},
  pages   = {6684--6707},
  doi     = {10.1287/mnsc.2023.4756}
}

@article{samuelson1958,
  author  = {Samuelson, Paul A.},
  title   = {An Exact Consumption-Loan Model of Interest with or without the Social Contrivance of Money},
  journal = {Journal of Political Economy},
  year    = {1958},
  volume  = {66},
  number  = {6},
  pages   = {467--482}
}

@article{tirole1985asset,
  author  = {Tirole, Jean},
  title   = {Asset Bubbles and Overlapping Generations},
  journal = {Econometrica},
  year    = {1985},
  volume  = {53},
  number  = {5},
  pages   = {1071--1100}
}

@article{weil1987confidence,
  author  = {Weil, Philippe},
  title   = {Confidence and the Real Value of Money in an Overlapping Generations Economy},
  journal = {The Quarterly Journal of Economics},
  year    = {1987},
  volume  = {102},
  number  = {1},
  pages   = {1--22}
}

@incollection{blanchardwatson1982,
  author    = {Blanchard, Olivier J. and Watson, Mark W.},
  title     = {Bubbles, Rational Expectations and Financial Markets},
  booktitle = {Crises in the Economic and Financial Structure},
  editor    = {Wachtel, Paul},
  publisher = {Lexington Books},
  year      = {1982}
}

@article{stiglitzweiss1981,
  author  = {Stiglitz, Joseph E. and Weiss, Andrew},
  title   = {Credit Rationing in Markets with Imperfect Information},
  journal = {The American Economic Review},
  year    = {1981},
  volume  = {71},
  number  = {3},
  pages   = {393--410}
}

@article{barlevy2014leverage,
  author  = {Barlevy, Gadi},
  title   = {A Leverage-Based Model of Speculative Bubbles},
  journal = {Journal of Economic Theory},
  year    = {2014},
  volume  = {153},
  pages   = {459--505},
  doi     = {10.1016/j.jet.2014.07.012}
}

@article{allenbarlevygale2022,
  author  = {Allen, Franklin and Barlevy, Gadi and Gale, Douglas},
  title   = {Asset Price Booms and Macroeconomic Policy: A Risk-Shifting Approach},
  journal = {American Economic Journal: Macroeconomics},
  year    = {2022},
  volume  = {14},
  number  = {2},
  pages   = {243--280},
  doi     = {10.1257/mac.20200041}
}

@article{allengale2000,
  author  = {Allen, Franklin and Gale, Douglas},
  title   = {Bubbles and Crises},
  journal = {The Economic Journal},
  year    = {2000},
  volume  = {110},
  number  = {460},
  pages   = {236--255}
}

@article{ikedaphan2016toxic,
  author  = {Ikeda, Daisuke and Phan, Toan},
  title   = {Toxic Asset Bubbles},
  journal = {Economic Theory},
  year    = {2016},
  volume  = {61},
  number  = {2},
  pages   = {241--271},
  doi     = {10.1007/s00199-015-0928-1}
}

@article{benguiphan2018pledgeability,
  author  = {Bengui, Julien and Phan, Toan},
  title   = {Asset Pledgeability and Endogenously Leveraged Bubbles},
  journal = {Journal of Economic Theory},
  year    = {2018},
  volume  = {177},
  pages   = {280--314},
  doi     = {10.1016/j.jet.2018.06.005}
}

@article{ikedaphan2019imbalances,
  author  = {Ikeda, Daisuke and Phan, Toan},
  title   = {Asset Bubbles and Global Imbalances},
  journal = {American Economic Journal: Macroeconomics},
  year    = {2019},
  volume  = {11},
  number  = {3},
  pages   = {209--251},
  doi     = {10.1257/mac.20140286}
}

@article{asriyanfornaromartinventura2021monetary,
  author  = {Asriyan, Vladimir and Fornaro, Luca and Mart{\'\i}n, Alberto and Ventura, Jaume},
  title   = {Monetary Policy for a Bubbly World},
  journal = {The Review of Economic Studies},
  year    = {2021},
  volume  = {88},
  number  = {3},
  pages   = {1418--1456},
  doi     = {10.1093/restud/rdaa045}
}

@article{farhitirole2012bubbly,
  author  = {Farhi, Emmanuel and Tirole, Jean},
  title   = {Bubbly Liquidity},
  journal = {The Review of Economic Studies},
  year    = {2012},
  volume  = {79},
  number  = {2},
  pages   = {678--706}
}

@article{martinventura2012growth,
  author  = {Mart{\'\i}n, Alberto and Ventura, Jaume},
  title   = {Economic Growth with Bubbles},
  journal = {American Economic Review},
  year    = {2012},
  volume  = {102},
  number  = {6},
  pages   = {3033--3058}
}

@article{martinventura2018guide,
  author  = {Mart{\'\i}n, Alberto and Ventura, Jaume},
  title   = {The Macroeconomics of Rational Bubbles: A User's Guide},
  journal = {Annual Review of Economics},
  year    = {2018},
  volume  = {10},
  pages   = {505--539},
  doi     = {10.1146/annurev-economics-080217-053534}
}

@article{endershakenes2021,
  author  = {Enders, Zeno and Hakenes, Hendrik},
  title   = {Market Depth, Leverage, and Speculative Bubbles},
  journal = {Journal of the European Economic Association},
  year    = {2021},
  volume  = {19},
  number  = {5},
  pages   = {2577--2621},
  doi     = {10.1093/jeea/jvab008}
}

@article{fukui2018asset,
  author  = {Fukui, Masao},
  title   = {Asset Quality Cycles},
  journal = {Journal of Monetary Economics},
  year    = {2018},
  volume  = {95},
  pages   = {97--108},
  doi     = {10.1016/j.jmoneco.2018.02.006}
}

@article{caramp2025sowing,
  title={Sowing the seeds of financial crises: Endogenous asset creation and adverse selection},
  author={Caramp, Nicolas},
  journal={Review of Economic Studies},
  volume={92},
  number={5},
  pages={2861--2892},
  year={2025},
  publisher={Oxford University Press UK}
}

@book{laffont2002theory,
  title={The theory of incentives: the principal-agent model},
  author={Laffont, Jean-Jacques and Martimort, David},
  year={2002},
  publisher={Princeton University Press}
}

@article{budish2025trust,
  author  = {Budish, Eric},
  title   = {Trust at Scale: The Economic Limits of Cryptocurrencies and Blockchains},
  journal = {The Quarterly Journal of Economics},
  year    = {2025},
  volume  = {140},
  number  = {1},
  pages   = {1--62},
  doi     = {10.1093/qje/qjae033}
}

@article{saleh2021blockchain,
  author  = {Saleh, Fahad},
  title   = {Blockchain without Waste: Proof-of-Stake},
  journal = {The Review of Financial Studies},
  year    = {2021},
  volume  = {34},
  number  = {3},
  pages   = {1156--1190},
  doi     = {10.1093/rfs/hhaa075}
}

@article{biais2019blockchain,
  author  = {Biais, Bruno and Bisi{\`e}re, Christophe and Bouvard, Matthieu and Casamatta, Catherine},
  title   = {The Blockchain Folk Theorem},
  journal = {The Review of Financial Studies},
  year    = {2019},
  volume  = {32},
  number  = {5},
  pages   = {1662--1715},
  doi     = {10.1093/rfs/hhy095}
}

@article{biais2023bitcoin,
  author  = {Biais, Bruno and Bisi{\`e}re, Christophe and Bouvard, Matthieu and Casamatta, Catherine and Menkveld, Albert J.},
  title   = {Equilibrium {Bitcoin} Pricing},
  journal = {The Journal of Finance},
  year    = {2023},
  volume  = {78},
  number  = {2},
  pages   = {967--1014},
  doi     = {10.1111/jofi.13206}
}

@article{garrattwallace2018bitcoin,
  author  = {Garratt, Rodney and Wallace, Neil},
  title   = {{Bitcoin} 1, {Bitcoin} 2, \ldots: An Experiment in Privately Issued Outside Monies},
  journal = {Economic Inquiry},
  year    = {2018},
  volume  = {56},
  number  = {3},
  pages   = {1887--1897},
  doi     = {10.1111/ecin.12569}
}

@article{howellniessneryermack2020,
  author  = {Howell, Sabrina T. and Niessner, Marina and Yermack, David},
  title   = {Initial Coin Offerings: Financing Growth with Cryptocurrency Token Sales},
  journal = {Review of Financial Studies},
  year    = {2020},
  volume  = {33},
  number  = {9},
  pages   = {3925--3974},
  doi     = {10.1093/rfs/hhz131}
}

@article{jensenmeckling1976,
  author  = {Jensen, Michael C. and Meckling, William H.},
  title   = {Theory of the Firm: Managerial Behavior, Agency Costs and Ownership Structure},
  journal = {Journal of Financial Economics},
  year    = {1976},
  volume  = {3},
  number  = {4},
  pages   = {305--360},
  doi     = {10.1016/0304-405X(76)90026-X}
}

@article{bravgompers2003,
  author  = {Brav, Alon and Gompers, Paul A.},
  title   = {The Role of Lockups in Initial Public Offerings},
  journal = {Review of Financial Studies},
  year    = {2003},
  volume  = {16},
  number  = {1},
  pages   = {1--29},
  doi     = {10.1093/rfs/16.1.0001}
}

@article{edmans2009,
  author  = {Edmans, Alex},
  title   = {Blockholder Trading, Market Efficiency, and Managerial Myopia},
  journal = {Journal of Finance},
  year    = {2009},
  volume  = {64},
  number  = {6},
  pages   = {2481--2513},
  doi     = {10.1111/j.1540-6261.2009.01508.x}
}

@article{ofekrichardson2003,
  author  = {Ofek, Eli and Richardson, Matthew},
  title   = {{DotCom} Mania: The Rise and Fall of Internet Stock Prices},
  journal = {Journal of Finance},
  year    = {2003},
  volume  = {58},
  number  = {3},
  pages   = {1113--1137},
  doi     = {10.1111/1540-6261.00560}
}

@article{hongscheinkmanxiong2006,
  author  = {Hong, Harrison and Scheinkman, Jos{\'e} and Xiong, Wei},
  title   = {Asset Float and Speculative Bubbles},
  journal = {Journal of Finance},
  year    = {2006},
  volume  = {61},
  number  = {3},
  pages   = {1073--1117},
  doi     = {10.1111/j.1540-6261.2006.00867.x}
}

\end{document}